\documentclass[article]{aa}
\usepackage[switch]{lineno}
\usepackage{natbib}
\bibpunct{(}{)}{;}{a}{}{,} 
\usepackage{bm}
\usepackage{multirow}
\usepackage{caption}
\usepackage{subcaption}
\usepackage{hyperref}
\usepackage{tikz}
\usepackage[frozencache,cachedir=.]{minted}

\title{\texttt{agnpy}: An open-source python package modelling the radiative processes of jetted active galactic nuclei}

\newcommand{\python}{\texttt{python}\xspace}
\newcommand{\numpy}{\texttt{NumPy}\xspace}
\newcommand{\scipy}{\texttt{SciPy}\xspace}
\newcommand{\astropy}{\texttt{astropy}\xspace}
\newcommand{\agnpy}{\texttt{agnpy}\xspace}
\newcommand{\sherpa}{\texttt{sherpa}\xspace}
\newcommand{\gammapy}{\texttt{Gammapy}\xspace}
\newcommand{\naima}{\texttt{naima}\xspace}
\newcommand{\jetset}{\texttt{jetset}\xspace}
\newcommand{\Beta}{\mathcal{B}}

\newcommand{\sedunits}{{\rm erg}\,{\rm cm}^{-2}\,{\rm s}^{-1}}
\newcommand{\diff}{\mathrm{d}}

\newcommand{\edensity}{\underline{n}'_{\rm e}}
\newcommand{\enumber}{\underline{N}'_{\rm e}}
\newcommand{\phdensity}{\underline{u}}

\newcommand*\patchAmsMathEnvironmentForLineno[1]{%
  \expandafter\let\csname old#1\expandafter\endcsname\csname #1\endcsname
  \expandafter\let\csname oldend#1\expandafter\endcsname\csname end#1\endcsname
  \renewenvironment{#1}%
     {\linenomath\csname old#1\endcsname}%
     {\csname oldend#1\endcsname\endlinenomath}}%
\newcommand*\patchBothAmsMathEnvironmentsForLineno[1]{%
  \patchAmsMathEnvironmentForLineno{#1}%
  \patchAmsMathEnvironmentForLineno{#1*}}%
\AtBeginDocument{%
\patchBothAmsMathEnvironmentsForLineno{equation}%
\patchBothAmsMathEnvironmentsForLineno{align}%
\patchBothAmsMathEnvironmentsForLineno{flalign}%
\patchBothAmsMathEnvironmentsForLineno{alignat}%
\patchBothAmsMathEnvironmentsForLineno{gather}%
\patchBothAmsMathEnvironmentsForLineno{multline}%
}

\author{
C.~Nigro\inst{\ref{inst1}}\thanks{corresponding author: \texttt{cosimo.nigro@ifae.es}} \and 
J.~Sitarek\inst{\ref{inst2}} \and
P.~Gliwny\inst{\ref{inst3}} \and 
D.~Sanchez\inst{\ref{inst4}} \and
A.~Tramacere\inst{\ref{inst5}} \and
M.~Craig\inst{\ref{inst6}}
}

\institute{Institut de F\'isica d’Altes Energies (IFAE), The Barcelona Institute of Science and Technology, Campus UAB, 08193 Bellaterra (Barcelona), Spain\label{inst1} \and 
University of Lodz, Faculty of Physics and Applied Informatics, Department of Astrophysics, 90-236 Lodz, Poland \label{inst2} \and
University of Lodz, Doctoral School of Exact and Natural Sciences, Banacha Street 12/16, 90-237 Lodz, Poland \label{inst3} \and 
Laboratoire d'Annecy de Physique des Particules, Univ. Grenoble Alpes, Univ. Savoie Mont Blanc, CNRS,  LAPP, 74000 Annecy, France \label{inst4} \and
Department of Astronomy, University of Geneva, ch. d'Ecogia 16, CH-1290 Versoix, Switzerland \label{inst5} \and
Department of Physics and Astronomy, Minnesota State University Moorhead, 1104 7th Ave S, Moorhead, MN 56563 \label{inst6}
}

\date{}

\abstract{
Modelling the broadband emission of jetted active galactic nuclei (AGN) constitutes one of the main research topics of extragalactic astrophysics in the multi-wavelength and multi-messenger domain.
}{
We present \agnpy, an open-source python package modelling the radiative processes of relativistic particles accelerated in the jets of AGN. The package includes classes describing the galaxy components responsible for line and thermal emission and it calculates the absorption due to $\gamma\gamma$ pair production on several photon fields. \agnpy aims to extend the effort of modelling and interpreting the emission of extragalactic sources to a wide number of astrophysicists.
}{
We present the package content and illustrate a few examples of applications of its functionalities. We validate the software by comparing its results against the literature and against other open-source software.
}{
We illustrate the utility of \agnpy in addressing the most common questions encountered while modelling the emission of jetted active galaxies. When comparing its results against the literature and other modelling tools adopting the same physical assumptions, we achieve an agreement within $10-30\%$.
}{
\agnpy represents one of the first systematic and validated collection of established radiative processes for jetted active galaxies in an open-source python package. We hope it will also stand among the first endeavours providing reproducible and transparent astrophysical software not only for data reduction and analysis, but also for modelling and interpretation. 
}


\begin{document}
\titlerunning{\texttt{agnpy}: an open-source python package modelling the radiative processes of jetted AGN}
\maketitle

\section{Introduction}
Active galactic nuclei (AGN) with jets, which are the brightest among the persistent extragalactic sources, present broadband emission covering the whole electromagnetic spectrum. Modelling this emission with the radiative processes of relativistic charged particles allows one to derive their non-thermal energy distributions, to infer their acceleration mechanism and site, and to investigate the jet composition (see \citealt{cerruti_2020} for a review centred on the radiative processes and \citealt{boettcher_2019} for a broader overview of the topic). The operation in the last two decades of space-borne and ground-based gamma-ray instruments has extended the spectral energy distributions (SEDs) of these sources with simultaneous data up to ${\rm TeV}$ energies \citep{sikora_2016}. Physical interpretation of their multi-wavelength (MWL) observations has traditionally been provided, even in the publications by the experiments and observatories gathering the data, by a few individuals with proprietary and private modelling software. The latter, though advancing the theoretical understanding of the field, is not accessible to the community at large and produces results difficult to reproduce or verify \citep{portgies_2018}. Moreover, the operation of the next generation of gamma-ray observatories such as the Cherenkov Telescope Array (CTA) and the larger sample of simultaneous complete SEDs they will facilitate \citep[][Chs. 8 and 12]{science_cta_book}, calls for the modelling and interpretation effort to be opened to a wider number of astrophysicists. For all these reasons, it is desirable to develop a new generation of open-source AGN modelling software.
\par
It has become increasingly common to build open-source astrophysical libraries in the ecosystem formed by the \numpy \citep{numpy}, \scipy \citep{scipy}, and \astropy \citep{astropy1, astropy2} libraries. The \astropy project gathers $\sim40$ packages built on these fundamental dependencies in a community of `affiliated packages' committed to create reusable and interfaceable astronomical python software\footnote{\href{https://www.astropy.org/affiliated/\#affiliated-package-list}{https://www.astropy.org/affiliated/\#affiliated-package-list}}. Notable among these are \gammapy \citep{gammapy}, for the analysis of high-energy astrophysical data, adopted as science analysis tool for CTA, and \naima \citep{naima}, the first python package modelling the non-thermal SED of astrophysical objects. Designed for sources with unbeamed radiation, \naima computes the radiative processes assuming comoving densities of interacting particles. It is therefore appropriate for interpreting the emission of galactic sources like supernova remnants, but inadequate when considering jetted extragalactic sources. In this case, relativistic transformations have to be applied on the interacting particle (or photon) densities, that are usually moving at relativistic speed with respect to each other. Additionally, the emission from the jet has to be boosted, accounting for its beaming in a cone inclined at an angle $\theta_{\rm s}$ with respect to the observer line of sight. \texttt{jetset} \citep{tramacere_2020} offers an open-source modelling framework for extragalactic and galactic jetted sources: blazars \citep{massaro_2006, tramacere_2009, tramacere_2011} and microquasars \citep{rodi_2021}. The \jetset core libraries compute the photon spectra due to leptonic (due to electrons and positrons) radiative processes and ${\rm p}{\rm p}$ interactions and evolve the particle distributions accounting for their acceleration and cooling. Written in \texttt{C}, they are wrapped with a high-level \python interface relying on \numpy and \astropy, connecting the framework to this environment. The high-level interface includes routines for MWL SED data handling and fitting using other \python data-analysis packages as backend.
\par
We created \agnpy with the objective of providing an open-source python package for jetted AGN modelling entirely built in the \numpy, \scipy, and \astropy ecosystem, interfaceable with data-analysis tools such as \gammapy and completing the science cases covered by \naima and \jetset. \agnpy gathers and implements the leptonic radiative processes most commonly considered in jetted AGN, its main  objective being modelling the broadband SED, from radio to gamma rays, produced by these sources. Expanding the package's scope beyond the non-thermal SED computation, we included additional classes describing the AGN components emitting thermal and line radiation. \agnpy is able to compute the absorption due to $\gamma\gamma$ pair production on the soft photon fields of the different AGN components and provides the absorption due to the extragalactic background light (EBL). Though not designed for time evolution, the package offers the possibility of a self-consistent constraint of the model parameters. Rather than solving the differential equation regulating the evolution of the electron energy distribution (EED), \agnpy assigns its maximum and break Lorentz factors using a simple parametrisation of the acceleration, escape, and radiation processes. Throughout all the physical processes described in the package, the angle of the jet axis to the observer line of sight, $\theta_{\rm s}$, is always considered among the model parameters, such that the use of \agnpy is not restricted to the blazar case ($\theta_{\rm s} \rightarrow 0$) but can also be extended to radio galaxies.
\par 
This article is structured as follows: in Sect.~\ref{sec:content} we describe the package implementation and content; in Sect.~\ref{sec:applications} we expose some of its basic functionalities and in Sect.~\ref{sec:crosschecks} we validate its results against the literature and other modelling software. In Sect.~\ref{sec:reproducibility} we discuss reproducibility and performance of our software. We gather our conclusion and thoughts for future expansions in Sect.~\ref{sec:conclusions}. The appendixes collect the mathematical derivations used in the code implementation, the benchmark model parameters used for the validation and some technical considerations.

\section{Package description}
\label{sec:content} 

\begin{figure*}
\centering
\includegraphics[scale=0.65]{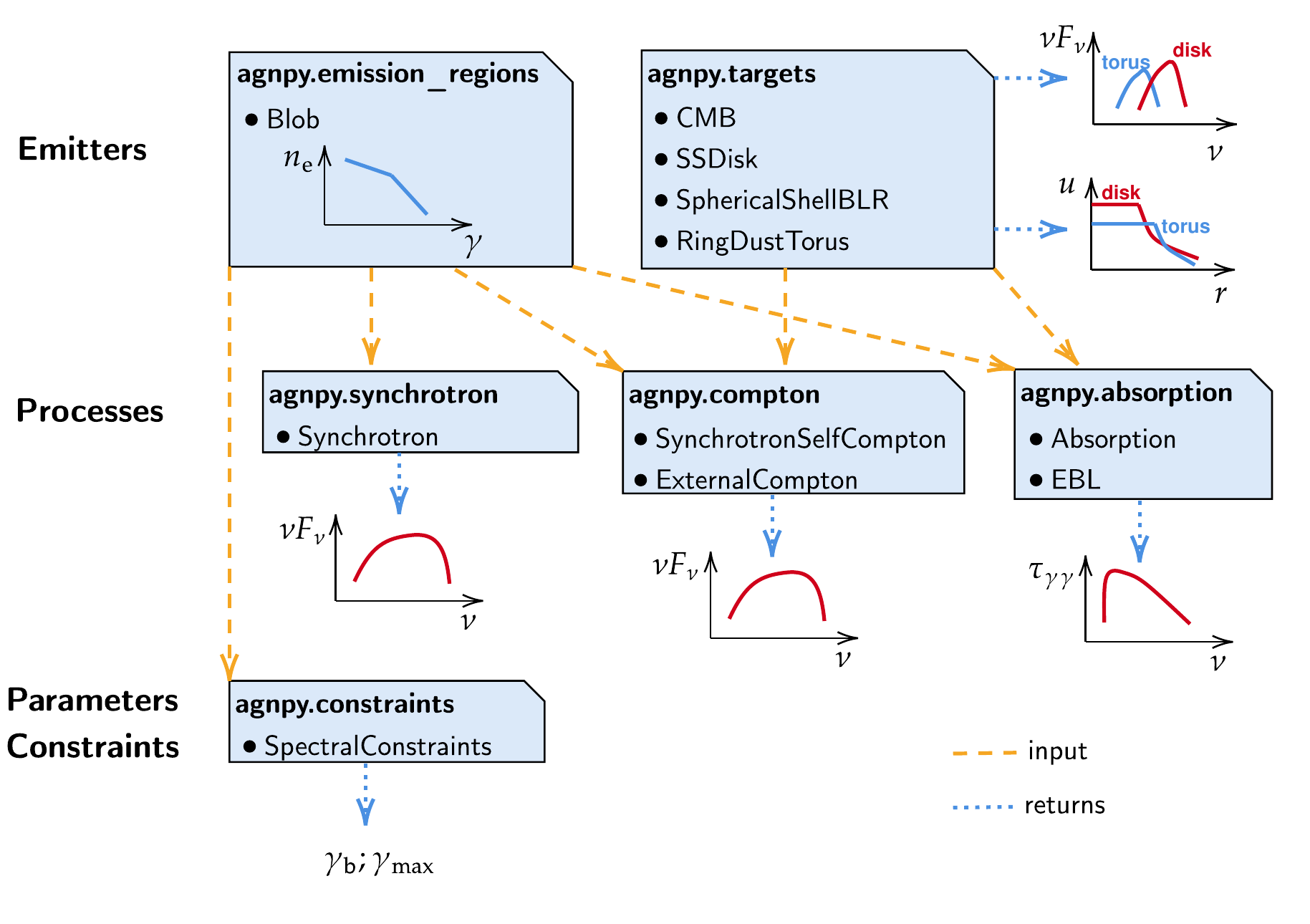}
\caption{Scheme representing \agnpy's modules, their relations and the computations they perform. \texttt{agnpy.emission\_regions}, defining the region responsible for the non-thermal radiation (containing the EED), is the initialising argument of the classes modelling radiative processes and computing their SEDs (\texttt{agnpy.synchrotron} and \texttt{agnpy.compton}). Thermal and line emitters, collected in \texttt{agnpy.targets}, are initialising arguments of the modules for the external Compton and the $\gamma\gamma$ absorption calculations (\texttt{agnpy.absorption}). The targets' thermal SEDs and energy densities ($u$) can also be computed. The emission region is passed to the class for self-consistent modelling (\texttt{agnpy.constraints}) and to the absorption class to compute the $\gamma\gamma$ absorption on synchrotron photons.}
\label{fig:package_scheme}
\end{figure*}

The basic formulae and notation used in the package follow \citet{dermer_menon_2009}. Assumptions and references employed for the specific radiative processes implementation are specified in what follows. The multidimensional integration needed to compute the emission of the electron distributions or of the photon fields (see App.~\ref{sec:formulae}) is performed with \numpy. \astropy provides all of the physical constants, handles unit transformations and distance calculations. The plotting utilities for visualisation of results rely on \texttt{matplotlib} \citep{matplotlib}. The code development is hosted on \texttt{GitHub}\footnote{\href{https://github.com/cosimoNigro/agnpy}{https://github.com/cosimoNigro/agnpy}} and its documentation on \texttt{readthedocs}\footnote{\href{https://agnpy.readthedocs.io/en/latest/index.html}{https://agnpy.readthedocs.io/en/latest/index.html}}. Each \agnpy release tagged in \texttt{GitHub} is archived on the Zenodo platform that offers a digital object identifier (DOI) for each release of the code. Version \texttt{0.1.3} \citep{agnpy_zenodo} was used for this publication. Like \gammapy and \naima, \agnpy is one of the packages affiliated with the \astropy project. The input quantities and the results of all the computations of \agnpy are provided as \astropy \texttt{Quantities} (or arrays of them, as in the case of a SED), allowing a seamless integration of other package using the same library upstream or downstream of the calculations performed with \texttt{agnpy}.
\par
In what follows we detail the content of the individual modules constituting the package. Fig.~\ref{fig:package_scheme} offers an overview of their relations and the computations they perform.

\subsection{Emitters}
\agnpy contains modules describing thermal and non-thermal emitters.
\par
\texttt{agnpy.emission\_regions} describes the emission regions responsible for the non-thermal radiative processes. The emission region contains the EED: the distribution of Lorentz factors, $\gamma'$, of the ${\rm e}^{\pm}$ accelerated in the source, $\edensity(\gamma')\,[{\rm cm}^{-3}]$ (primed quantities refer to the reference frame comoving with the blob, see App.~\ref{sec:formulae} for more details on the notation). At the moment the only non-thermal emission region implemented is a spherical plasmoid streaming along the jet (blob). The electron distribution is considered homogeneous and isotropic in the blob's rest frame.
\par
\texttt{agnpy.targets} describes line and thermal emitters providing the photon fields target for inverse Compton scattering or $\gamma\gamma$ absorption. The line and thermal emitters modelled in the package are: the cosmic microwave background (CMB); a monochromatic point-like source behind the jet (to be used mostly for consistency checks, see Sect.~\ref{sec:crosschecks}); a \cite{shakura} accretion disc, modelled as a geometrically thin disc whose emission is parameterised following \citet{dermer_2002} and \citet{dermer_2009}; a broad line region (BLR) simplified to an infinitesimally thin sphere reprocessing a fraction of the disc radiation in a monochromatic (line) emission \citep[see Sect~3.5 in][and references therein]{finke_2016}\footnote{All the emission lines of the stratified BLR model of \citet{finke_2016} are made available.}; a dust torus (DT), simplified to a ring reprocessing a fraction of the disc radiation in a single-temperature black body (BB) emission \citep[see Sect.~3.6 of][]{finke_2016}. The thermal emission of some of the targets can also be computed: one can evaluate the multi-temperature BB SED produced in optical-UV by the accretion disc or the IR-peaking BB SED representing the DT emission. These BB SEDs approximate the actual thermal emission and are conventionally adopted in radio-to-gamma-ray MWL modelling to check the flux level of the thermal components against the dominant non-thermal ones \citep[see e.g.][]{ghisellini_2009}. They are inadequate for precise modelling of the SED of those sources in which the thermal components can be directly observed (in particular non-jetted AGN). The energy density, $u\,[{\rm erg}\,{\rm cm}^{-3}]$, of every target can also be computed as a function of the distance from the central black hole (BH), $r$, both in the reference frame comoving with the blob and in the one comoving with the centre of the galaxy (see Fig.~\ref{fig:package_scheme}).

\subsection{Radiative processes}
\label{sec:processes}
The package contains a module for each leptonic radiative process \citep[see][for canonical references]{blumenthal_gould_1970, rybicki_lightman_1986, dermer_menon_2009}. All the formulae used in the radiative processes calculations are provided in Apps.~\ref{sec:synchrotron_radiation}-\ref{sec:absorption}. Here we describe the fundamental physical assumptions considered in their implementation.
\par
\texttt{agnpy.synchrotron} computes synchrotron spectra (with self-absorption), following \citet{finke_2008}. It is assumed that the electron distribution is immersed in a large-scale random magnetic field \citep{crusius_1986}.
\par 
\texttt{agnpy.compton} models the inverse Compton (IC) scattering either of the synchrotron photons produced by the same accelerated electrons, `synchrotron self-Compton' (SSC), or of the soft photon fields produced by the AGN line and thermal emitters or by the CMB, `external Compton' (EC). In both cases, the full Compton cross section, including the Klein-Nishina regime, is considered. SSC spectra are computed as in \citet{finke_2008}, with the target synchrotron radiation assumed uniform in the blob \citep{jones_1968}. EC spectra are computed according to \citet{dermer_2009} and \citet{finke_2016}. The electron distribution is transformed to a frame comoving with the target photon field \citep[following the approach of][]{gkm_2001} and then convolved with the energy distribution of the latter and the Compton cross section. The head-on approximation of the cross section is used: it is assumed that the scattered photons have roughly the same direction of the scattering electrons \citep[as in][]{dermer_1993, dermer_2002}.
\par 
\texttt{agnpy.absorption} allows one to compute the absorption due to $\gamma\gamma$ pair production \citep{gould_1967} on the soft photon fields of the line and thermal emitters, as well as on the synchrotron photons produced in the emission region. Concerning $\gamma\gamma$ pair production with the EBL, absorption values computed according to the models of \citet{franceschini_2008}, \citet{finke_2010}, and \citet{dominguez_2011} are available in the package resources. They are interpolated to provide an energy- and redshift-dependent opacity function.

\subsection{Parameter constraints}
The package does not include any routine for the solution of the differential equation describing the EED time evolution \citep[see e.g.][Eq.~7]{cerruti_2020}. Nonetheless it contains an additional module, \texttt{agnpy.constraints}, constraining its spectral parameters such that the modelling can be considered self-consistent. Three time scales (computed in the reference frame of the emission region and therefore indicated with primed quantities, see notation in App.~\ref{sec:formulae}) are considered in the code: acceleration, cooling and dynamical or ballistic. The acceleration time scale of electrons with energy $E'$ is defined by the acceleration parameter $\xi$ \citep[see e.g.][]{bednarek_2007} as
\begin{equation}
    t'_{\rm acc} = E' / \left(\frac{\diff E'}{\diff t'}\right)_{\rm acc}, \text{ where } \left(\frac{\diff E'}{\diff t'}\right)_{\rm acc} = \xi c E' / R_{\rm L}, 
    \label{eq:t_acc}
\end{equation}
where $c$ is the speed of light and $R_{\rm L}$ is the Larmor radius of the electron. 
The cooling time scale due to energy losses via synchrotron radiation or inverse Compton in the Thomson regime can be computed as \citep{blumenthal_gould_1970}
\begin{equation}
    t'_{\rm cool} = E' / \left(\frac{\diff E'}{\diff t'}\right)_{\rm cool}, \text{ where } \left(\frac{\diff E'}{\diff t'}\right)_{\rm cool} = \frac{4}{3} \sigma_T c u' \gamma'^2,
    \label{eq:t_cool}
\end{equation}
where $\sigma_T$ is the Thomson cross section and $u'$ is the energy density of the magnetic field (for synchrotron cooling), or the radiation field (for inverse Compton cooling in the Thomson regime), computed in the reference frame of the emission region. We note that, due to relativistic beaming, the radiation density of the photon field of emitters external to the blob has to be transformed to its comoving reference frame (see  App.~\ref{sec:energy_densities} for these transformations). Finally, the ballistic or dynamic time scale is determined from the blob size via
\begin{equation}
t'_{\mathrm{bal}} = R'_{\rm b} / c. 
\label{eq:t_bal}
\end{equation}
This time scale determines the escape time of the radiation from the blob. As the electrons are confined in the blob by the magnetic field they are naturally not directly affected by such an escape condition. Nevertheless Eq.~(\ref{eq:t_bal}) also determines the time scale in which the blob crosses a distance corresponding to its size. In other words, in the reference frame comoving with the blob, the ambient matter traverses a distance corresponding to the radius of the emission region. Both the escape of the radiation and the change of the blob's surrounding conditions can cause breaks in the spectrum. It should be noticed that the exemplary broken power law implicitly assumed will not accurately describe the evolving electron distribution in all cases. In the case of hard injection spectra, fast cooling can result in an additional pile-up in the EED \citep[see e.g.][]{sh04}

\section{Applications}
\label{sec:applications}
In the following section we illustrate some applications of the package functionalities, performing typical estimations necessary for modelling the emission of jetted AGN.

\subsection{Computing the SED for a radiative process}
\begin{figure*}
    \begin{subfigure}[t]{0.49\textwidth}
        \vspace{0.2cm}
        \begin{minted}[fontsize=\footnotesize]{python}
import numpy as np
import astropy.units as u
from agnpy.emission_regions import Blob
from agnpy.synchrotron import Synchrotron
from agnpy.utils.plot import plot_sed
import matplotlib.pyplot as plt
            
# emission region and radiative process
blob = Blob()
synch = Synchrotron(blob)

# compute the SED over an array of frequencies
nu = np.logspace(8, 23) * u.Hz
sed = synch.sed_flux(nu)

# plot it
plot_sed(nu, sed, label="synchrotron")
plt.show()
        \end{minted}
        \caption{Example script using \agnpy to compute and display a synchrotron SED.}
        \label{fig:synchrotron_snippet}
    \end{subfigure}
    \hfill
    \begin{subfigure}[t]{0.49\textwidth}
        \centering
        \vspace{0pt}
        \includegraphics[width=\textwidth]{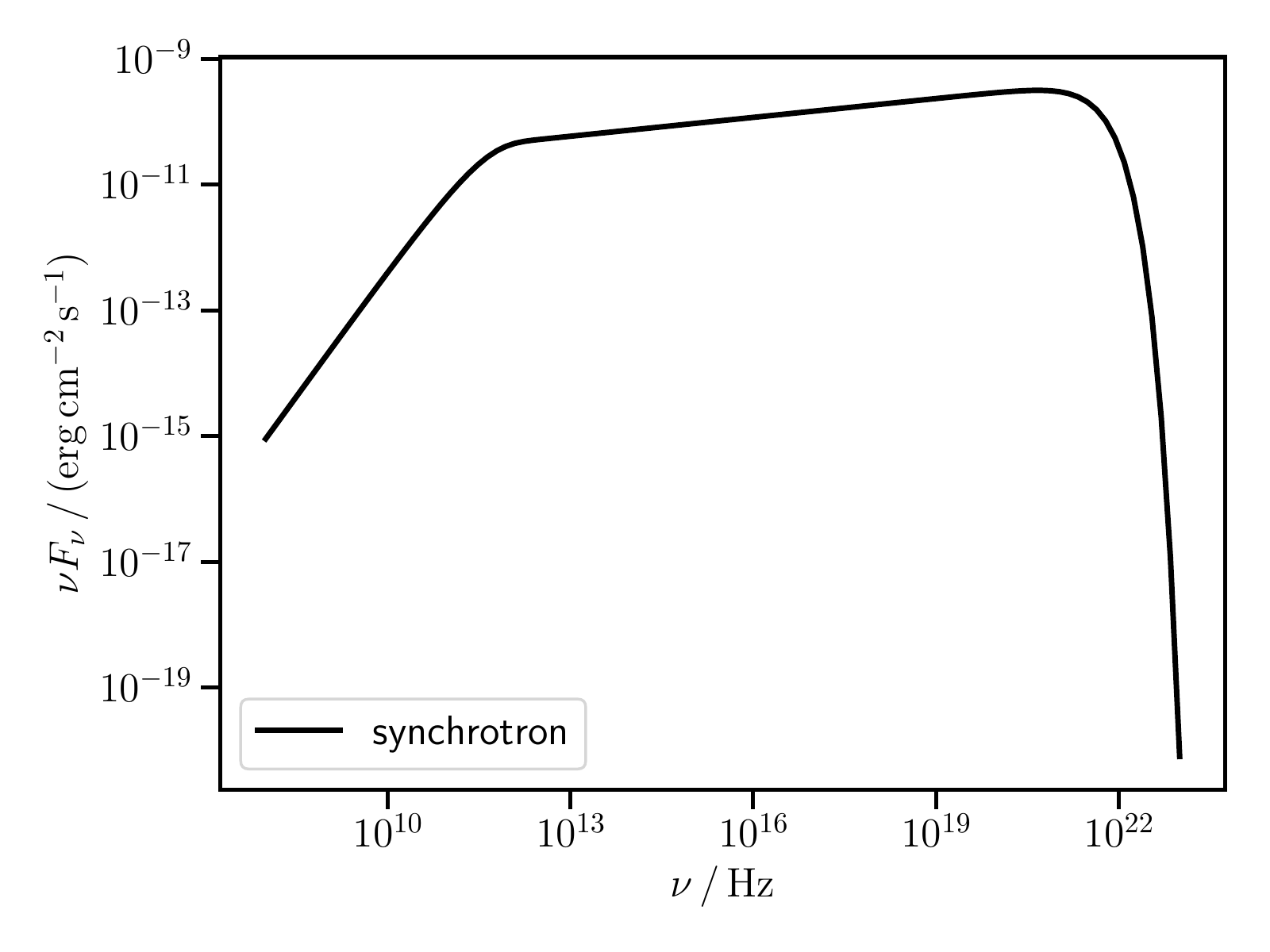}
        \caption{Output of the example script.}
        \label{fig:synchrotron_sed}
    \end{subfigure}
    \hfill
\caption{SED due to synchrotron radiation computed with \agnpy. Default parameters for the emission region are given in Table~\ref{tab:blob_parameters} (SSC model column); fluxes and frequencies are given in the observer reference frame.}
\label{fig:synchrotron}
\end{figure*}

As stated in the introduction, the main objective of the package is the computation of the broadband SED, from radio to gamma rays, due to the radiative process in jetted AGN. 
In the code snippet in Fig.~\ref{fig:synchrotron_snippet} we illustrate how, with few lines of \python, one can evaluate the SED due to synchrotron radiation of a power law of electrons (Fig.~\ref{fig:synchrotron_sed}). SED computations for further radiative processes are illustrated in Sect.~\ref{subsec:fitting}, \ref{subsec:ssc_validation} and \ref{subsec:ec_validation}.

\subsection{Emission and densities of target photon fields}
\label{subsec:u_targets}
\begin{figure}
\resizebox{\hsize}{!}{\includegraphics{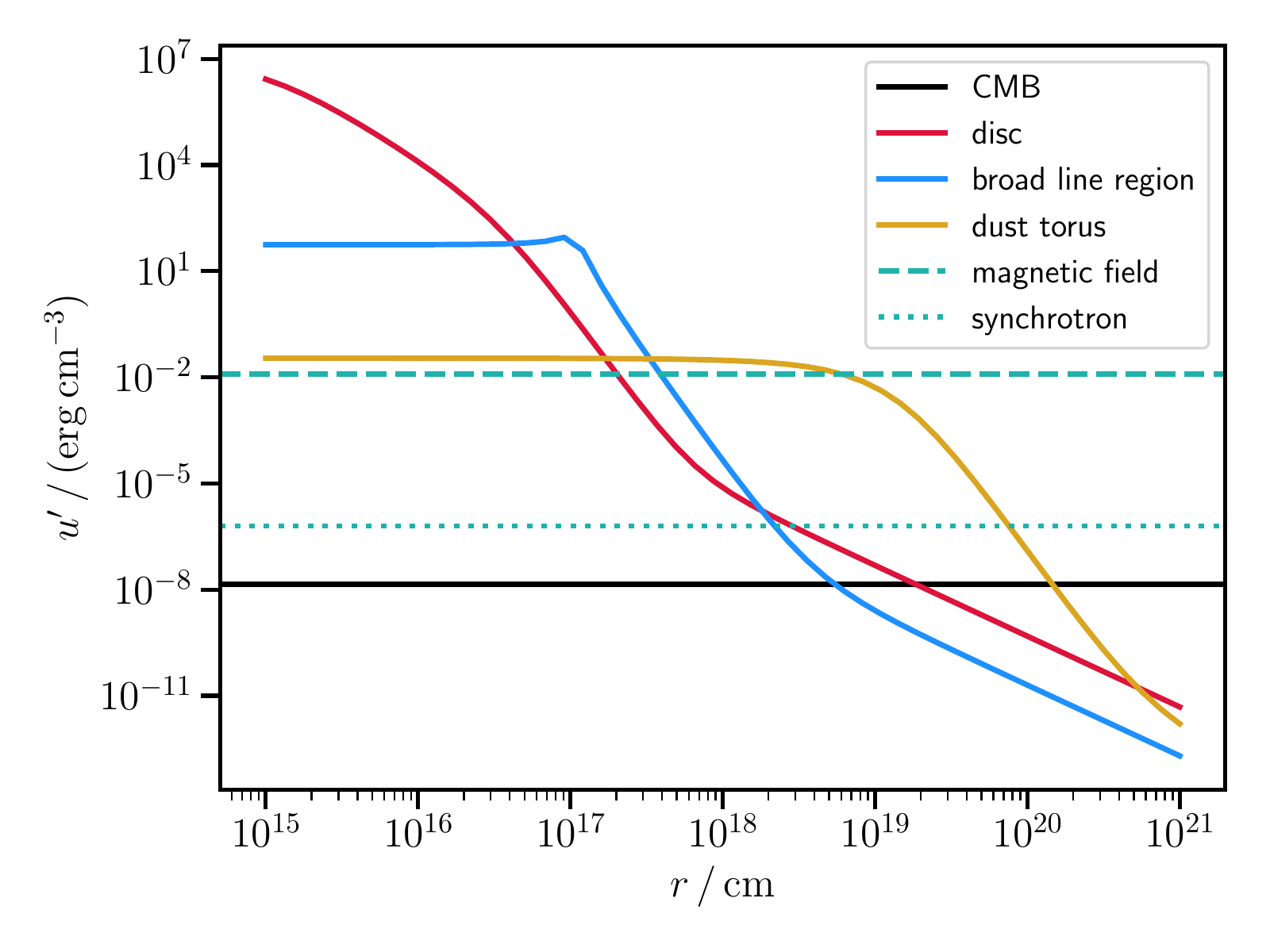}}
\caption{Energy densities of several photon fields: accretion disc (red solid line), BLR (blue solid line), DT (yellow solid line), CMB (black solid line), as a function of the distance from the central BH, computed with \agnpy. The energy density of synchrotron radiation (green dotted line) and of the magnetic field (green dashed line) are plotted for comparison. The emitters parameters are specified in Table~\ref{tab:target_parameters}. A BLR emitting the ${\rm Ly \alpha}$ line is considered. The energy densities are expressed in the reference frame comoving with the second blob in Table~\ref{tab:blob_parameters} (EC model).}
\label{fig:u_targets_example}
\end{figure}

When considering the inverse Compton scenario in presence of different photon targets, one might want to determine which one produces the highest energy density $u\,[{\rm erg}\,{\rm cm}^{-3}]$ (see App.~\ref{sec:energy_densities} for a definition, Sect.~3 of \citealt{ghisellini_2009} and Sect.~8 of \citealt{finke_2016} for a more complete discussion). In the Thomson regime, for example, the higher energy density will be the one dominating the electrons' Compton cooling, Eq.~(\ref{eq:t_cool}), as long as the radiation can be considered quasi-isotropic. \agnpy estimates the energy densities of line and thermal emitters as a function of the distance from the central black hole (BH) along the jet axis, $u(r)$, in the reference frames comoving with the galaxy or the blob. Figure~\ref{fig:u_targets_example} illustrates, as an example, the energy densities of the photon fields generated by the CMB, a Shakura-Sunyaev disc, a spherical shell BLR emitting the ${\rm Ly\alpha}$ line, and a ring DT, whose parameters are given in Table~\ref{tab:target_parameters}. The energy densities are computed in the reference frame comoving with the emission region (EC model column in Table~\ref{tab:blob_parameters}). The energy density provided by the magnetic field of the blob, $U_{B} = B^2 / (8\pi)$, and by the synchrotron radiation are plotted for comparison\footnote{We have assumed that the radius of the blob and the magnetic field do not change as the distance from the BH increases. Normally the blob size would increase and the magnetic field intensity decrease.}. We observe that the energy densities of the AGN thermal and line emitters drop as we move at a distance larger than their typical sizes (for target parameters refer to Table~\ref{tab:target_parameters}). For distances even larger than the dust torus radius $r > 10^{20}\,{\rm cm} \approx 32\,{\rm pc}$ the only relevant external photon field is the CMB, though, in the particular case realised in Fig.~\ref{fig:u_targets_example}, the synchrotron radiation will be the one dominating the cooling.

\subsection{Absorption}

\begin{figure}
\resizebox{\hsize}{!}{\includegraphics{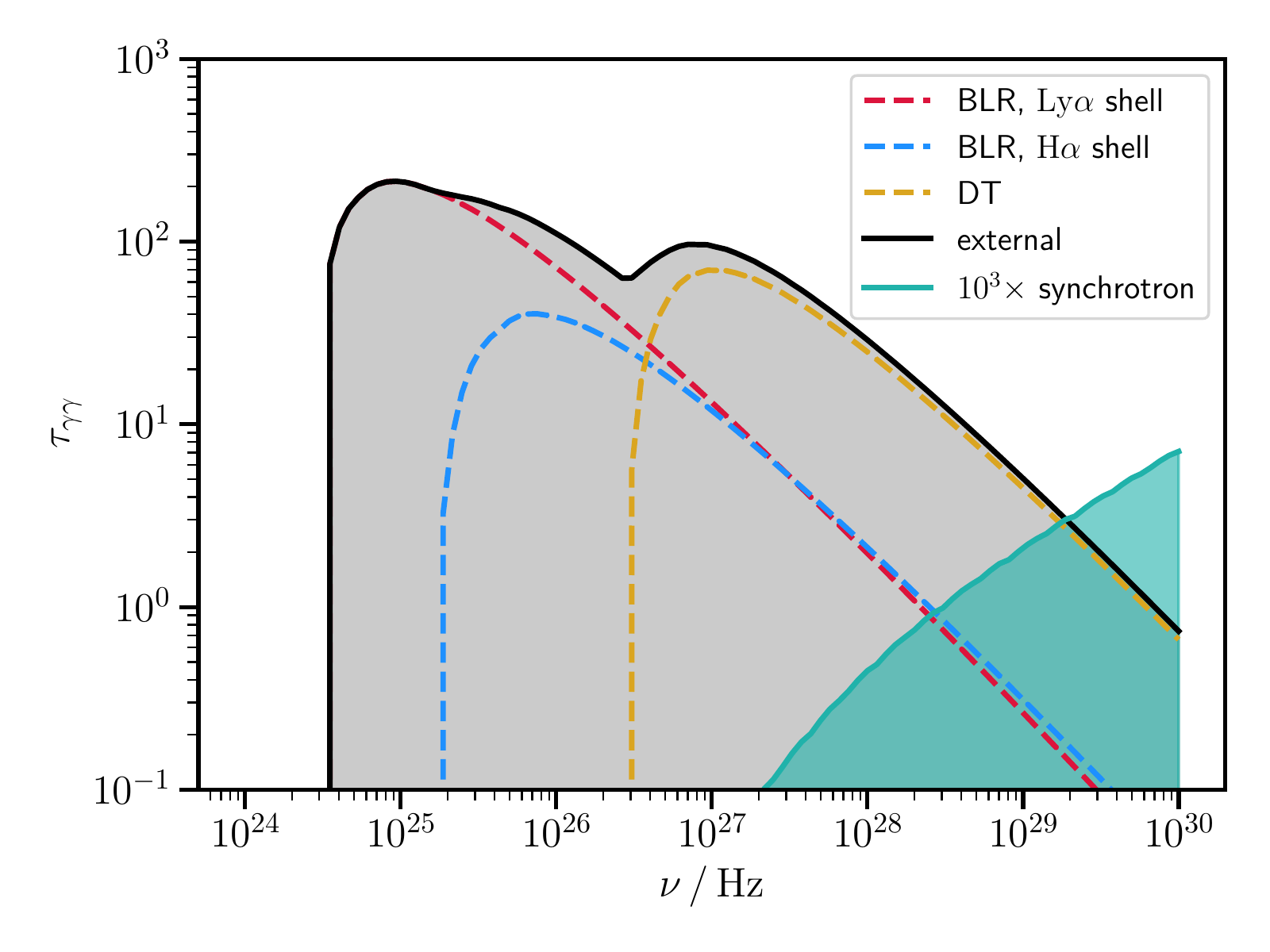}}
\caption{Total $\gamma\gamma$ opacity (black solid line and grey shaded area) produced by photon field external to the blob: BLR (red and blue dashed lines) and DT (yellow dashed line). The BLR is assumed to be composed of two concentric shells one emitting the ${\rm Ly\alpha}$ (red) and the other the ${\rm H \alpha}$ line (blue). Line and thermal emitters parameters are specified in Table~\ref{tab:target_parameters}. The green line and shaded area represent the opacity due to the synchrotron photons produced by the second blob in Table~\ref{tab:blob_parameters} (EC model). The latter is multiplied by $10^3$ to be comparable with the opacity due to the external photon fields. The blob is placed well within the BLR, at $r=1.1 \times 10^{16}\,{\rm cm} = 0.1 \times R_{\rm Ly\alpha}$. Frequencies are given in the observer reference frame.}
\label{fig:absorption_example}
\end{figure}

High-energy photons might be absorbed via $\gamma\gamma$ pair production by the very same soft photon fields that constitute the target for Compton scattering, or by the EBL on their path to Earth. \agnpy estimates the $\gamma\gamma$ opacity, $\tau_{\gamma\gamma}$, due to the soft photon fields of synchrotron radiation, of the BLR and the DT, and of a monochromatic point source behind the jet (though the latter is used only for cross-checks, as illustrated in Sect.~\ref{sec:crosschecks}). As an illustrative example, Fig.~\ref{fig:absorption_example} displays the total $\gamma\gamma$ opacity, computed with \agnpy, due to photon fields internal and external to the blob. For the internal photon fields, the opacity due to synchrotron radiation of the second blob in Table~\ref{tab:blob_parameters} (EC model) is represented, multiplied by a factor $10^3$ to be comparable with the opacity due to the external photon fields. For the latter, a DT and a BLR constituted of two concentrical shells, one emitting the ${\rm Ly \alpha}$ and the other the ${\rm H \alpha}$ line, are considered. The luminosity and radius of the ${\rm H \alpha}$ shell are obtained from those of the ${\rm Ly \alpha}$, following the scaling of the stratified BLR model in \citet{finke_2016}. Parameters of line and thermal emitters are given in Table~\ref{tab:target_parameters}. The blob is placed within the BLR, at a distance $r=1.1 \times 10^{16}\,{\rm cm} = 0.1 \times R_{\rm Ly\alpha}$. We note that internal and external absorption correspond to different attenuation factors of the final photon spectrum (see Sect.~\ref{sec:absorption}). It should be noticed that in the case of high opacity for high-energy photons, the secondary ${\rm e}^\pm$ pairs can produce further IC emission in the ambient radiation field \citep[see e.g.][]{bk95, bl95}. This cascading effect is not calculated by \agnpy.
\par
The package resources additionally include tables of the absorption value, $\exp(-\tau_{\gamma\gamma})$, computed according to the EBL models of \citet{franceschini_2008}, \citet{finke_2010}, and \citet{dominguez_2011}. These are interpolated and made available to the user to be evaluated, for a given redshift, over an array of frequencies.

\subsection{Constraining the Lorentz factor}
The module for constraining the spectral parameters provides a number of functions that compute the Lorentz factors of the electrons for which the different time scales listed in Eqs.~(\ref{eq:t_acc}) - (\ref{eq:t_bal}) match. Typically, a maximum Lorentz factor, $\gamma'_{\max}$, is obtained from the comparison between a given cooling and acceleration time scales. A break Lorentz factor, $\gamma'_{\rm b}$, corresponding to a change in slope in the EED, is instead obtained equating the cooling and the ballistic time scale. A practical use case would be to compute the $\gamma'_{\max}$ values for all the radiative processes considered in a particular scenario and then choose the lowest one as the most constraining. For example, considering the blob for the EC model in Table~\ref{tab:blob_parameters}, and the synchrotron, SSC, and EC on DT radiative processes, we obtain: $\gamma'_{\rm max,\,synch} = 1.6 \times 10^8$, $\gamma'_{\rm max,\,SSC} = 2.2 \times 10^{10}$, and $\gamma'_{\rm max,\,EC} = 2.4 \times 10^8$. We would then choose $\gamma'_{\rm max,\,synch}$ as a constraint on the maximum energy. The process responsible for the limit on the maximum energy of the electrons will likely produce a break in the electron spectrum connected with the change of the conditions inside the blob if the corresponding $\gamma'_{\rm b}$ is within the $[\gamma'_{\min}, \gamma'_{\max}]$ range, as it is in this case: $\gamma'_{\rm b,\,synch} = 7.4 \times 10^3$. Both spectral parameters are in fact in rough agreement with those chosen for the broken power-law EED in the EC model in the table.

\subsection{Fitting the SED of blazars}
\label{subsec:fitting}
\agnpy is a package for numerical modelling; routines for data handling and fitting are, therefore, outside of its scope. Moreover, such functionalities are available in many other packages and re-implementing them will, in addition to creating duplication, break the modular approach we discussed in the introduction. To ease the interface between \agnpy and other high-level data-analysis package we expose, for each radiative process, a function computing the SED that depends on the frequency and on all the model parameters simultaneously (i.e. $\nu F_{\nu}(\nu; \bm{\Theta})$, where $\bm{\Theta}$ is a vector of the model parameters one wants to determine through some statistical procedure). Such functions can be easily wrapped by any fitting routine, leaving substantial freedom to the user to create an optimal fit model from any arbitrary combination of emission regions and radiative processes. Once the model is defined, users can choose which parameters of the model have to be fixed and which one left free, depending on the completeness of their data set and on information already available on the source. We provide, in the online documentation tutorials and in the code associated with this publication, two examples using both \gammapy \citep{gammapy} version \texttt{0.18.2} \citep{gammapy_version} and \sherpa \citep{sherpa_1, sherpa_2} version \texttt{v4.14.0} to wrap \agnpy's functions to fit the observed MWL SED of the blazars Mrk~421 (a BL Lac) and PKS~1510-089 (a flat spectrum radio quasar, FSRQ). While the low energy continuum of both source types is interpreted as due to synchrotron radiation of electrons, the high-energy one is commonly explained as due to SSC for BL Lacs and to EC (with a possible additional contribution from SSC process) for FSRQs. These two science cases offer us the possibility to test the fitting procedure with most of the radiative processes implemented in the package. 

\subsubsection{Mrk~421}

\begin{figure}
\resizebox{\hsize}{!}{\includegraphics{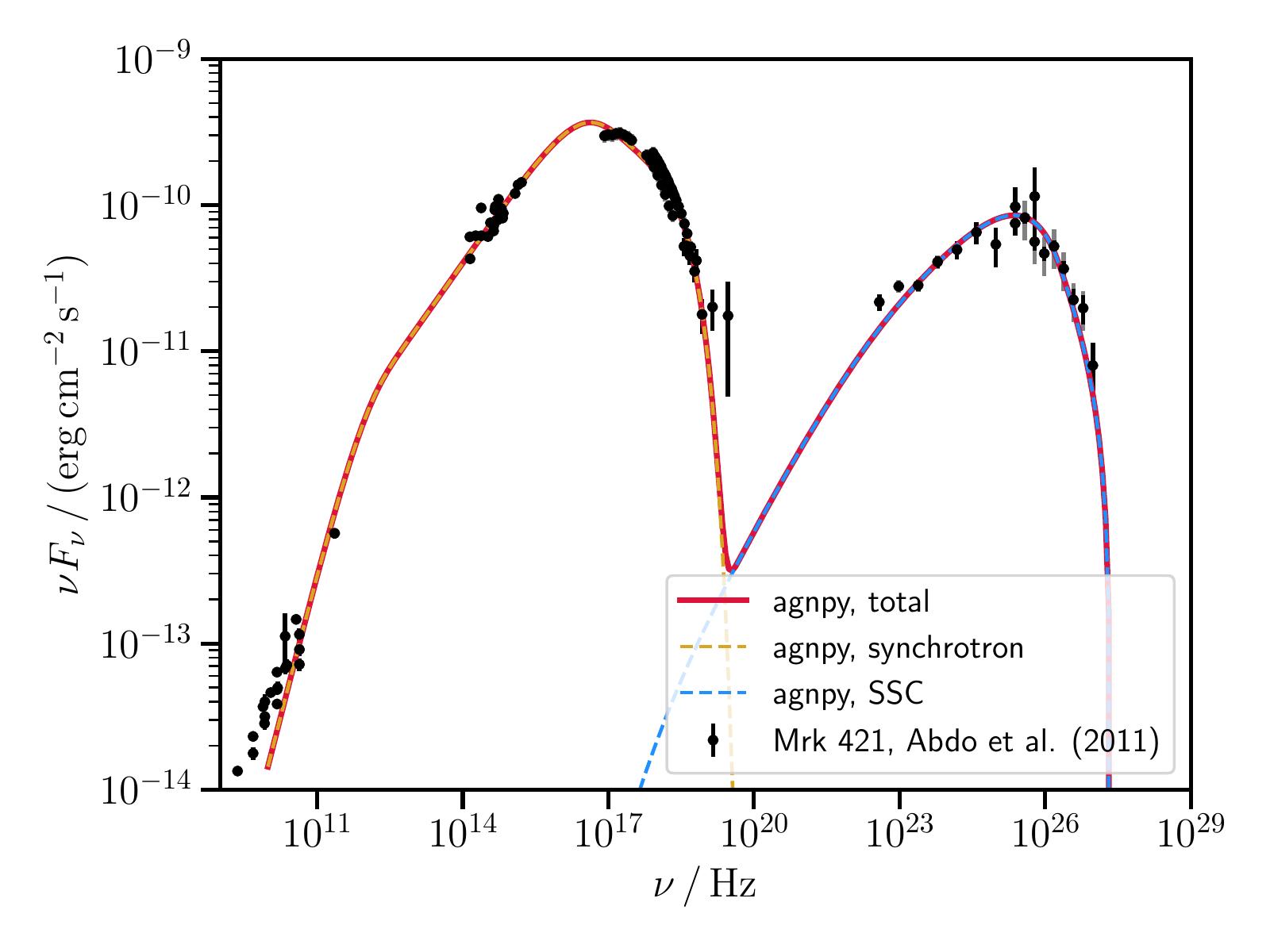}}
\caption{MWL SED of Mrk~421 (black points) from \citet{abdo_2011} fitted wrapping \agnpy with \gammapy. The red solid line represents the best fit model, the coloured dashed lines its individual components: yellow for synchrotron, blue for SSC. Fluxes and frequencies are given in the observer reference frame.}
\label{fig:mrk421_fit}
\end{figure}

We modelled the radio-to-gamma-ray broadband SED of Mrk~421 observed in 2009 by the MWL campaign described in \citet{abdo_2011} with the synchrotron and SSC radiation of a single blob accelerating a broken power-law EED. The model is described by nine parameters, six for the EED ($k_{\rm e}$, $p_1$, $p_2$, $\gamma'_{\rm b}$, $\gamma'_{\rm min}$, and $\gamma'_{\rm max}$; see the parametrisation in App.~\ref{sec:electron_densities}), and three for the blob ($R'_{\rm b}$, $B$, and $\delta_{\rm D}$). We fixed the minimum and maximum Lorentz factors of the EED, $\gamma'_{\rm min}$ and $\gamma'_{\rm max}$, and constrained the radius of the blob, $R'_{\rm b}$, with the observed variability time scale $t_{\rm var} = 1\,{\rm d}$ and the Doppler factor $\delta_D$, via $R'_{\rm b} = c t_{\rm var} \delta_{\rm D} / (1 + z)$. To limit the span of the parameters space, we fitted the $\log_{10}$ of the parameters whose range is expected to cover several orders of magnitudes, such as the electron normalisation, $k_{\rm e}$, the break Lorentz factors of the electron distribution, $\gamma'_{\rm b}$, and the magnetic field, $B$. Both for \gammapy and \sherpa we minimised a $\chi^2$ statistic. For \sherpa we selected the Levenberg-Marquardt minimisation algorithm \citep{levenberg_1944, marquardt_1963}, while for \gammapy we used \texttt{iminuit} \citep{iminuit}, a python wrapper for the \texttt{minuit2}\footnote{\href{https://root.cern.ch/root/htmldoc/guides/minuit2/Minuit2.html}{https://root.cern.ch/root/htmldoc/guides/minuit2/Minuit2.html}} library. The SED points below $10^{11}\,{\rm Hz}$ were not considered in the fit as they represent radio flux measurements with a larger integration region and are contaminated by the extended jet emission. We assumed $30\%$ systematic uncertainties on very-high-energy gamma-ray flux measurements ($E \geq 100\,{\rm GeV}$, \citealt{aleksic_2016}), $10\%$ on the high-energy ones ($0.1\,{\rm GeV} \leq E < 100\,{\rm GeV}$, \citealt{abdo_2011}), $10\%$ in the X-ray band \citep{madsen_2017}, and a conservative $5\%$ on the low-energy fluxes, from the UV to the radio band \citep{bohlin_2014}. The values of the model parameters obtained with the different software are reported in Table~\ref{tab:best_fit_params}. The best-fit model obtained with \gammapy, illustrated in Fig.~\ref{fig:mrk421_fit} along with its individual components, returned a statistics $\chi^2 = 271.2$ with $80$ degrees of freedom. We observe a good agreement between the fitted model and the data over the entire energy range. We refrain from providing uncertainties on the best-fit values due to the simplistic treatment of the flux points uncertainties: the systematic uncertainties we considered represent an educated guess from the literature and we also neglected any correlation between flux points obtained by the same instrument. We remark that the objective of this exercise is to illustrate the seamless interface of analysis and modelling tools and not to perform a statistical analysis nor draw any scientific conclusion from the modelling.

\subsubsection{PKS~1510-089}

\begin{figure}
\resizebox{\hsize}{!}{\includegraphics{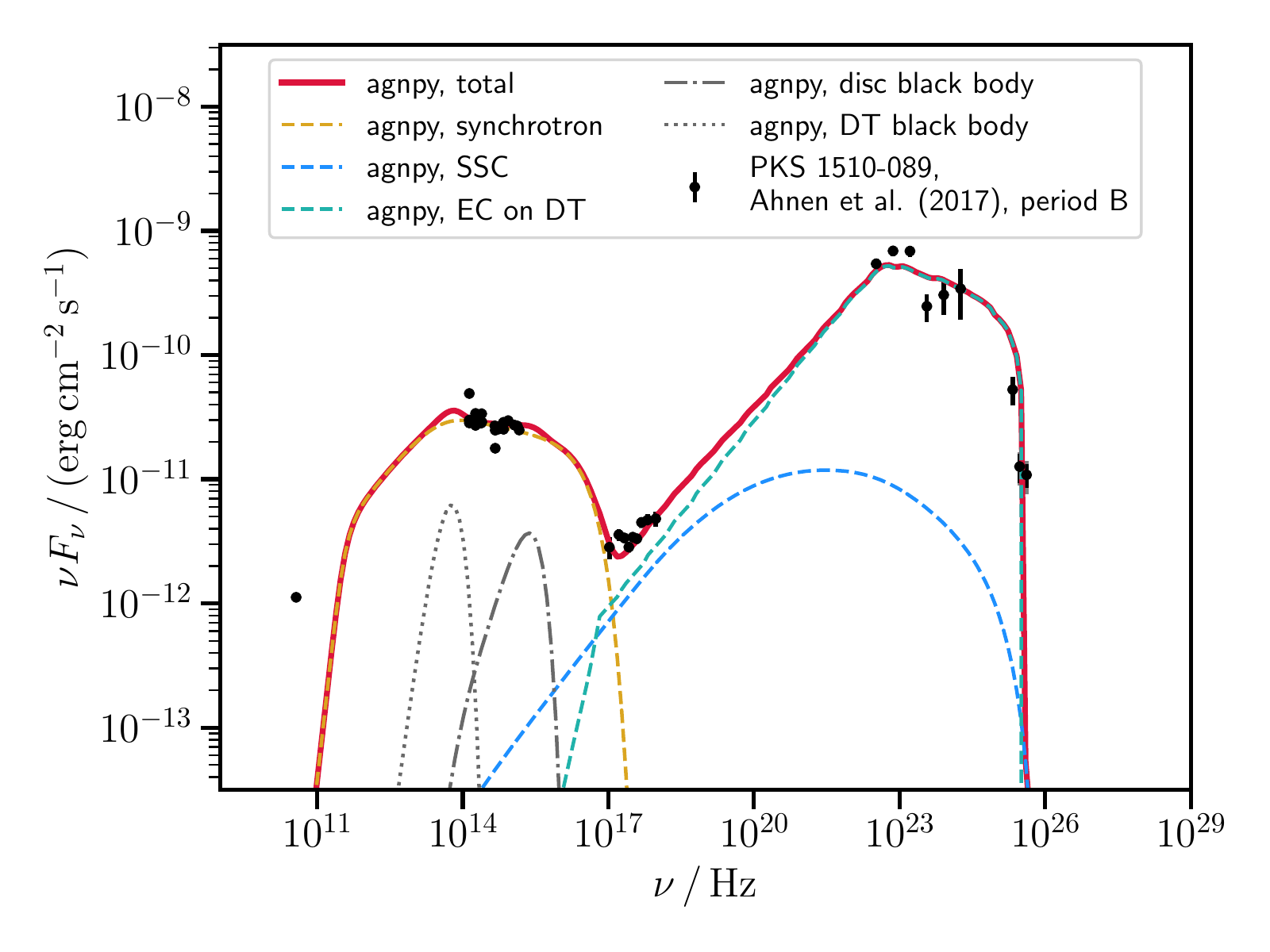}}
\caption{MWL SED of PKS~1510-089 (black points) from \citet{ahnen_2017} fitted wrapping \agnpy with \gammapy. The red solid line represents the best fit model, coloured dashed lines represent its non-thermal components: yellow for synchrotron, blue for SSC, green for EC on DT; grey lines represent its thermal components: dot-dashed for the multi-temperature BB of the accretion disc, dotted for the BB of the DT. Fluxes and frequencies are given in the observer reference frame.}
\label{fig:pks1510_fit}
\end{figure}

The MWL SED of PKS~1510-089, characteristic of a high optical and gamma-ray state observed in 2015 \citep{ahnen_2017} was modelled with the synchrotron, SSC and EC radiation of a single blob accelerating a broken power-law EED. We included, in the total flux model, the disc and the DT thermal emissions since for FSRQs, while often outshined by synchrotron emission, they are still non-negligible. We considered the DT photon field as target for the EC and neglected any $\gamma\gamma$ absorption on photon fields external to the blob. Independent studies on the location of the emission region indeed finds it to be outside the BLR, without any evidence of absorption of the gamma-ray spectrum on its photon field \citep{costamante_2018, meyer_2019}. For this reason we set the emission region location at a distance $r = 6 \times 10^{17}\,{\rm cm}$, outside the BLR \citep[following][]{ahnen_2017}. We accounted for self-absorption of the synchrotron radiation. For the $\chi^2$ minimisation we used the same settings described for Mrk~421. In this case the number of model parameters is larger than those of the simple synchrotron and SSC model, as they include the parameters of the disc and DT thermal components. We fixed all the parameters related to the thermal emitters: the disc luminosity $L_{\rm disc}$ and accretion efficiency $\eta$, the mass of its BH $M_{\rm BH}$, its inner and outer radii $R_{\rm in (out)}$. We considered a torus with radius $R_{\rm DT}$ reprocessing a fraction $\xi_{\rm DT}$ of the disc efficiency and radiating at temperature $T_{\rm DT}$. The values adopted, listed in Table~\ref{tab:fixed_params}, were obtained from previous modelling of the source \citep{aleksic_2014, ahnen_2017} following the scaling relations in \citet{ghisellini_2009}. As before, we fixed $\gamma'_{\rm min}$ and $\gamma'_{\rm max}$ of the EED, but due to the lower number of flux measurements compared to the Mrk~421 case, we were not able to fit simultaneously the EED parameters along with $\delta_{\rm D}$ and $B$ and therefore chose to fix the value of $\delta_{\rm D} = 25$. The blob size was again constrained by the time variability, fixed to $t_{\rm var} = 0.5\,{\rm d}$. As for Mrk~421, spectral points below $10^{11}\,{\rm Hz}$ were not included in the fit and we assumed for the flux points the same systematic uncertainties previously specified. The value of the model parameters obtained with the different software are reported in Table~\ref{tab:best_fit_params}. The best fit model obtained with \gammapy, illustrated in Fig.~\ref{fig:pks1510_fit} along with its thermal and non-thermal individual components, returns a statistics $\chi^2 = 230.5$ with $36$ degrees of freedom. Again we refrain from providing errors on the parameters.

\begin{table}
\caption{Parameters of Mrk~421 and PKS~1510-089 models from the examples wrapping \agnpy with \gammapy and \sherpa.}
\label{tab:fit_parameters}  

\begin{subtable}{1\linewidth}
\centering
\begin{tabular}{rrr|rr}        
\hline\hline
& \multicolumn{2}{c|}{Mrk~421} & \multicolumn{2}{c}{PKS~1510-089} \\
parameter & \gammapy & \sherpa & \gammapy & \sherpa \\ 
\hline
$\log_{10}(\frac{k_{\rm e}}{{\rm cm}^{-3}})$ & $-7.89$ & $-7.89$ & $-2.06$ & $-2.05$ \\
$p_1$ & $2.06$ & $2.06$ & $2.00$ & $2.00$ \\
$p_2$ & $3.54$ & $3.54$ & $3.16$ & $3.16$ \\
$\log_{10}(\gamma'_{\rm b})$ & $4.99$ & $4.99$ & $3.01$ & $3.01$ \\
$\log_{10}(B / {\rm G})$ & $-1.33$ & $-1.33$ & $-0.42$ & $-0.42$ \\
$\delta_{\rm D}$ & $19.74$ & $19.76$ & -- & -- \\
\hline                  
$\chi^2/{\rm d.o.f.}$ & $271.2 / 80$ & $271.2 / 80$ & $230.5 / 36$ & $230.5 / 36$ \\ 
\hline
\hline
\end{tabular}
\caption{Best-fit parameters.}
\label{tab:best_fit_params}
\end{subtable}

\hspace{\fill}

\bigskip
\begin{subtable}{1\linewidth}
\centering
\begin{tabular}{rr|r}          
\hline\hline
parameter & Mrk421 & PKS~1510-089 \\
\hline
$\delta_{\rm D}$ & -- & $25$ \\
$\gamma'_{\min}$ & $500$ & $1$ \\
$\gamma'_{\max}$ & $10^6$ & $3 \times 10^4$ \\
$R_{\rm b}\,/\,{\rm cm}$ & $5.3\times10^{16}$ & $2.4\times10^{16}$ \\
$\theta_{\rm s}$ & $2.90^{\circ}$ & $2.22^{\circ}$ \\
$r / {\rm cm}$ & -- & $6 \times 10^{17}$ \\
$L_{\rm disc} / ({\rm erg\,s}^{-1})$ & -- &  $6.7 \times 10^{45}$ \\
$\eta$ & -- & $1/12$ \\
$M_{\rm BH} / M_{\odot}$ & -- & $5.71 \times 10^{7}$ \\
$R_{\rm in} / R_{\rm g}$ & -- & $6$ \\
$R_{\rm out} / R_{\rm g}$ & -- & $10^4$ \\
$\xi_{\rm DT}$ & -- & $0.6$ \\
$R_{\rm DT}\,/\,{\rm cm}$ & -- & $6.5 \times 10^{18}$ \\
$T_{\rm DT}\,/\,{\rm K}$ & -- & $10^3$ \\
$z$ & $0.0308$ & $0.361$\\
\hline
\hline
\end{tabular}
\caption{Fixed parameters.}
\label{tab:fixed_params}
\end{subtable}

\end{table}

\section{Validation}
\label{sec:crosschecks}
In order to validate the calculations performed by \agnpy we compared them against bibliographic references and against \jetset (version \texttt{1.2.0rc10}). Moreover, we implemented internal consistency checks that examine the compatibility of different implementations of a similar scenario. One example is provided by the processes dependent on the distance between the emission region (the blob) and the central BH: EC and $\gamma\gamma$ absorption on AGN photon fields. The photon field produced by any emitter centred at the BH position can be approximated, for distances much larger than its size, $r \gg R_{\rm emitter}$, with the photon field produced by a point source (with the same luminosity) at the BH position. For very large distances, we can therefore verify the consistency of the EC spectrum (or the $\gamma\gamma$ opacity) generated by a given line or thermal emitter with the EC spectrum (or the opacity) generated by a point-source approximating it. All the comparisons illustrated in this section are embedded in the test system part of the package continuous integration (CI) cycle. Before a change is merged in the master branch of the code repository, these benchmark SEDs are automatically generated and it is numerically assessed that their deviations from the references are within a certain accuracy margin. 

\subsection{Synchrotron and synchrotron self-Compton}
\label{subsec:ssc_validation}

\begin{figure*}
\centering
\includegraphics[width=17cm]{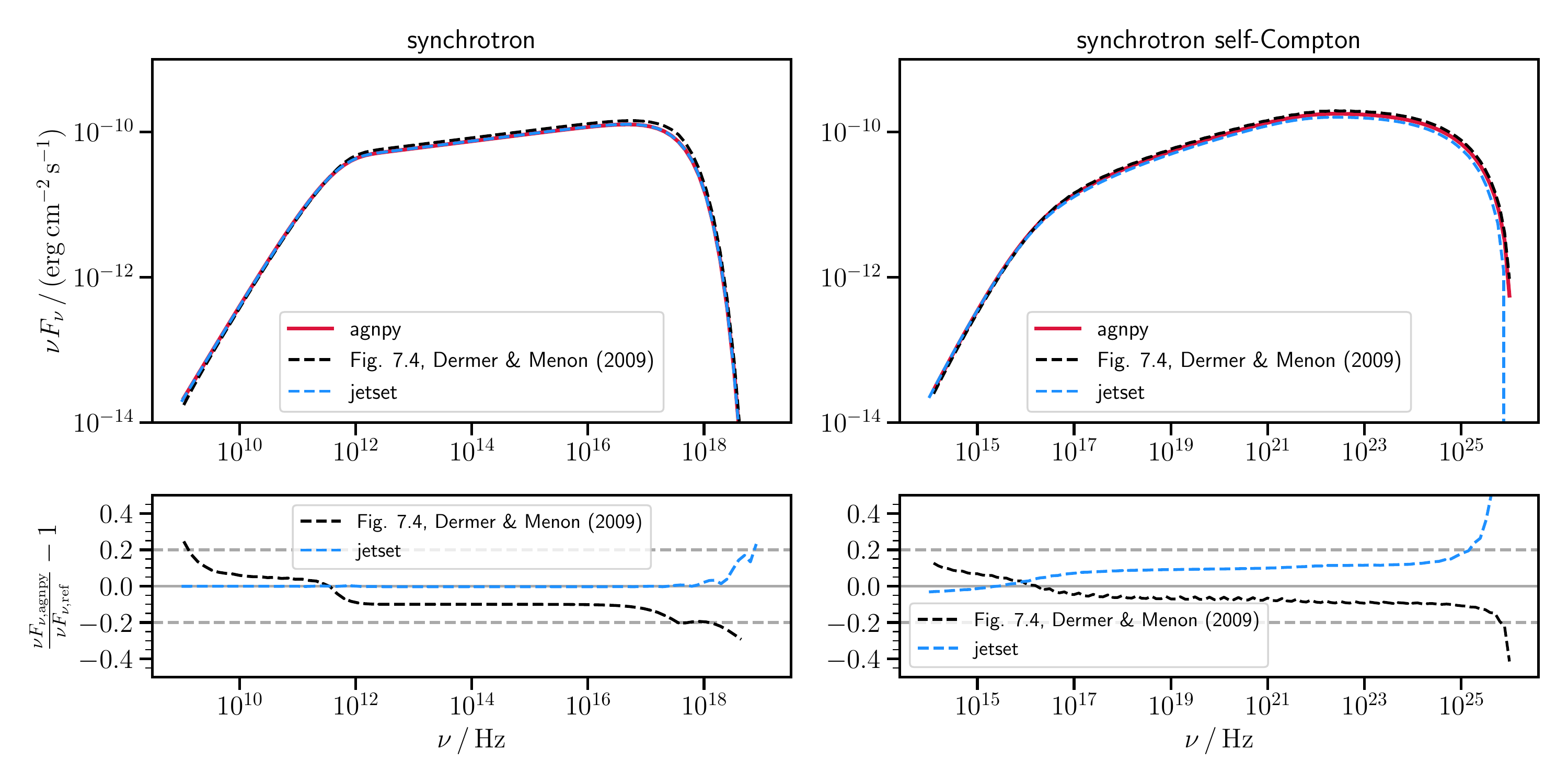}
\caption{Comparison of the synchrotron (left panels) and SSC (right panels) radiation computed with \agnpy (red solid line) and jetset (blue dashed line) and taken from Fig.7.4 of \citet{dermer_menon_2009} (black dashed line). The bottom panels display the deviation of \agnpy from the bibliographic reference or \jetset. Fluxes and frequencies are given in the observer reference frame.}
\label{fig:ssc_crosscheck}
\end{figure*}

To cross-check the synchrotron and SSC radiation calculation we reproduced with \agnpy one of the models in Fig.~7.4 of \citet{dermer_menon_2009} (whose parameters are given under the SSC model column of Table~\ref{tab:blob_parameters}). We used \jetset, which for these processes relies on the same assumptions as \agnpy (homogeneous magnetic field and homogeneous synchrotron radiation), to produce the SEDs for the same model parameters. Figure~\ref{fig:ssc_crosscheck} illustrates the comparison between the reference SEDs and \agnpy. In the bottom panels the deviation of \agnpy from a given reference is represented with the quantity $\frac{\nu F_{\nu, {\rm agnpy}}}{\nu F_{\nu, {\rm ref.}}} - 1$. We observe an absolute deviation for the synchrotron and SSC processes well within $20\%$ both from \citet{dermer_menon_2009} and from \jetset, for most of the frequency range considered.

\subsection{External Compton}
\label{subsec:ec_validation}

\begin{figure*}
\centering
    \includegraphics[width=17cm]{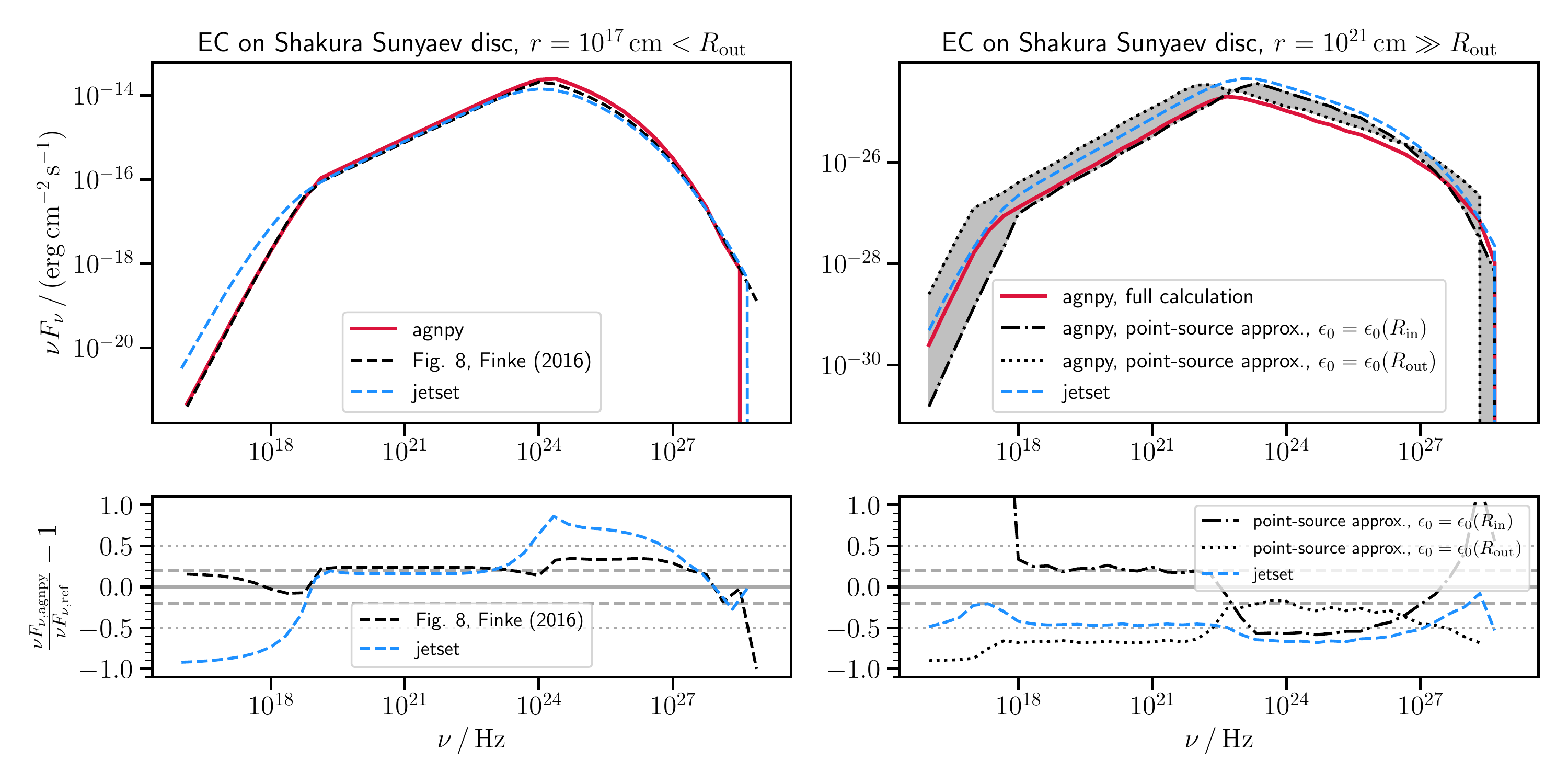}
    \caption{Left: comparison, for a distance smaller than the emitter size, of the EC scattering on the disc photon field computed with \agnpy (red solid line) and \jetset (dashed blue line) against the results in \citet{finke_2016} (black dashed line). Right: comparison, for a distance much larger than the emitter size, of the SED obtained with \agnpy considering the full disc energy density (solid red line) or approximating it with a monochromatic point-source (black line) with the same photon energy at the inner (dashed line) or outer (dotted line) disc radius. The SED computed by \jetset at the same distance is displayed with the dashed blue line. The bottom panels display the deviation of \agnpy from the reference (or the approximation). Fluxes and frequencies are given in the observer reference frame.}
    \label{fig:ec_disc_cosscheck}
\end{figure*}

\begin{figure*}
\centering
    \includegraphics[width=17cm]{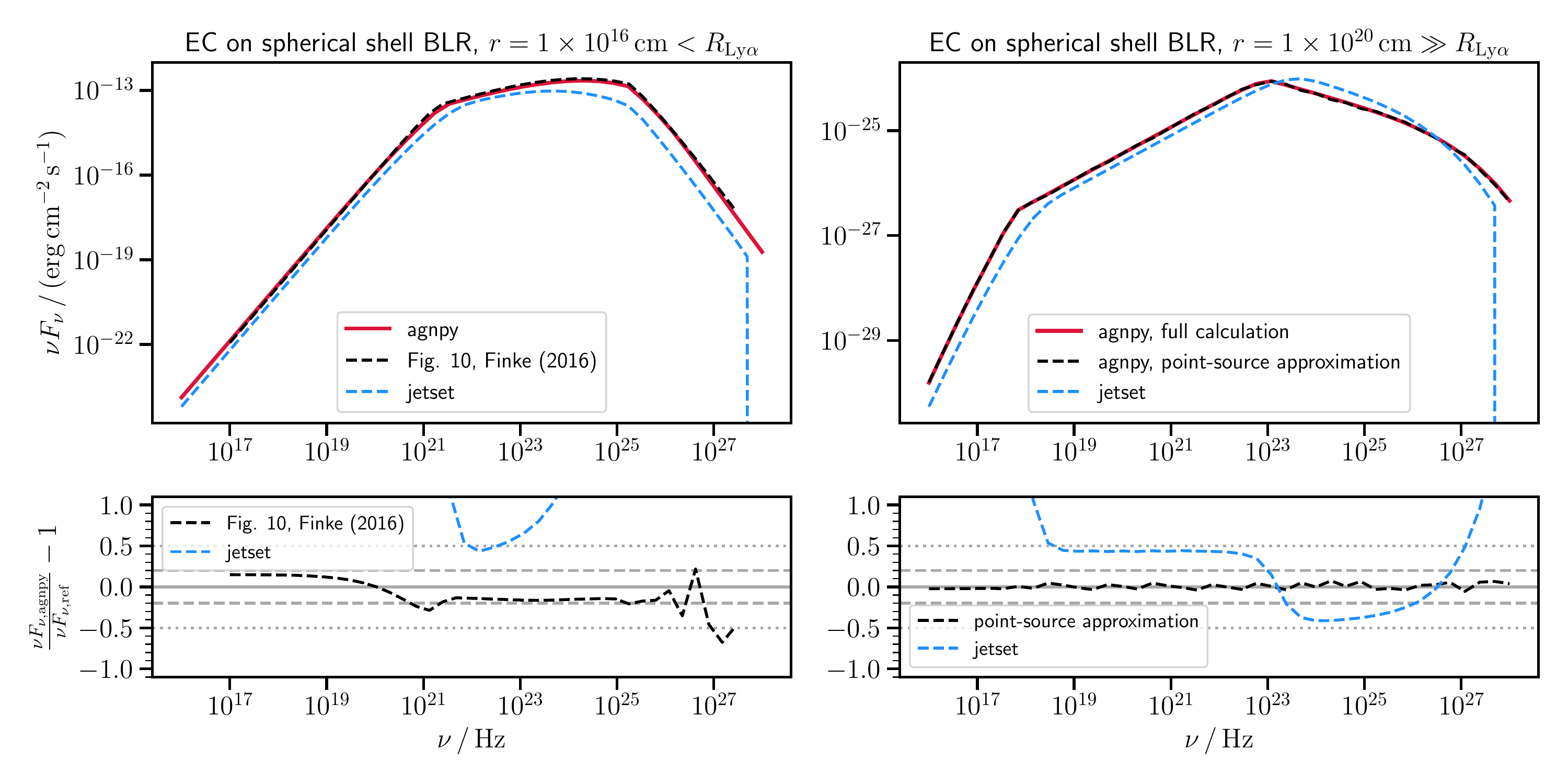}
    \caption{Same as in Fig.~\ref{fig:ec_disc_cosscheck}, but considering EC scattering on the BLR photon field.}
    \label{fig:ec_blr_crosscheck}
\end{figure*}

\begin{figure*}
\centering
    \includegraphics[width=17cm]{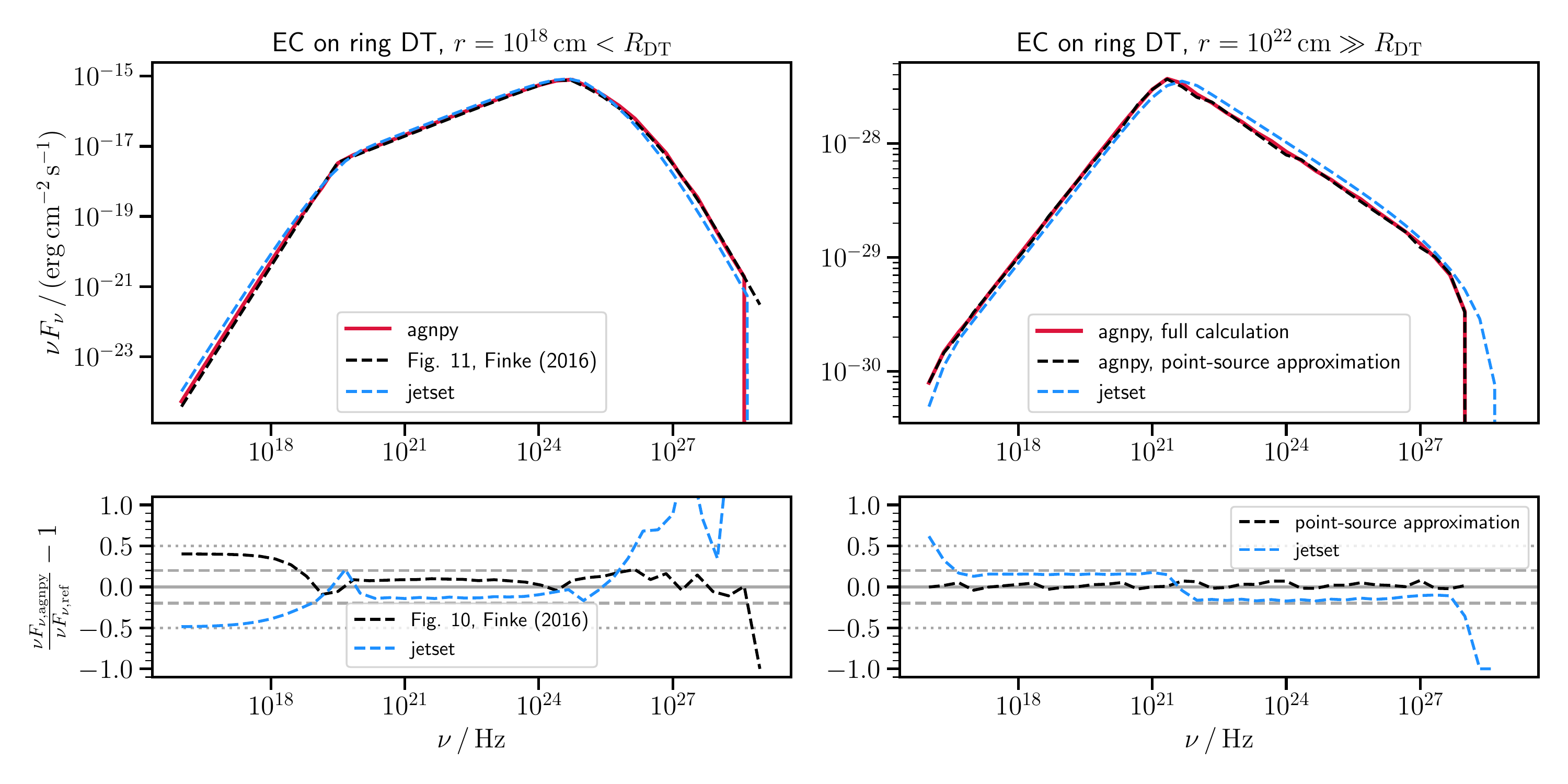}
    \caption{Same as in Fig.~\ref{fig:ec_disc_cosscheck} and Fig.~\ref{fig:ec_blr_crosscheck}, but considering EC scattering on the DT photon field.}
    \label{fig:ec_dt_crosscheck}
\end{figure*}

\begin{figure}
\centering
\resizebox{\hsize}{!}{\includegraphics{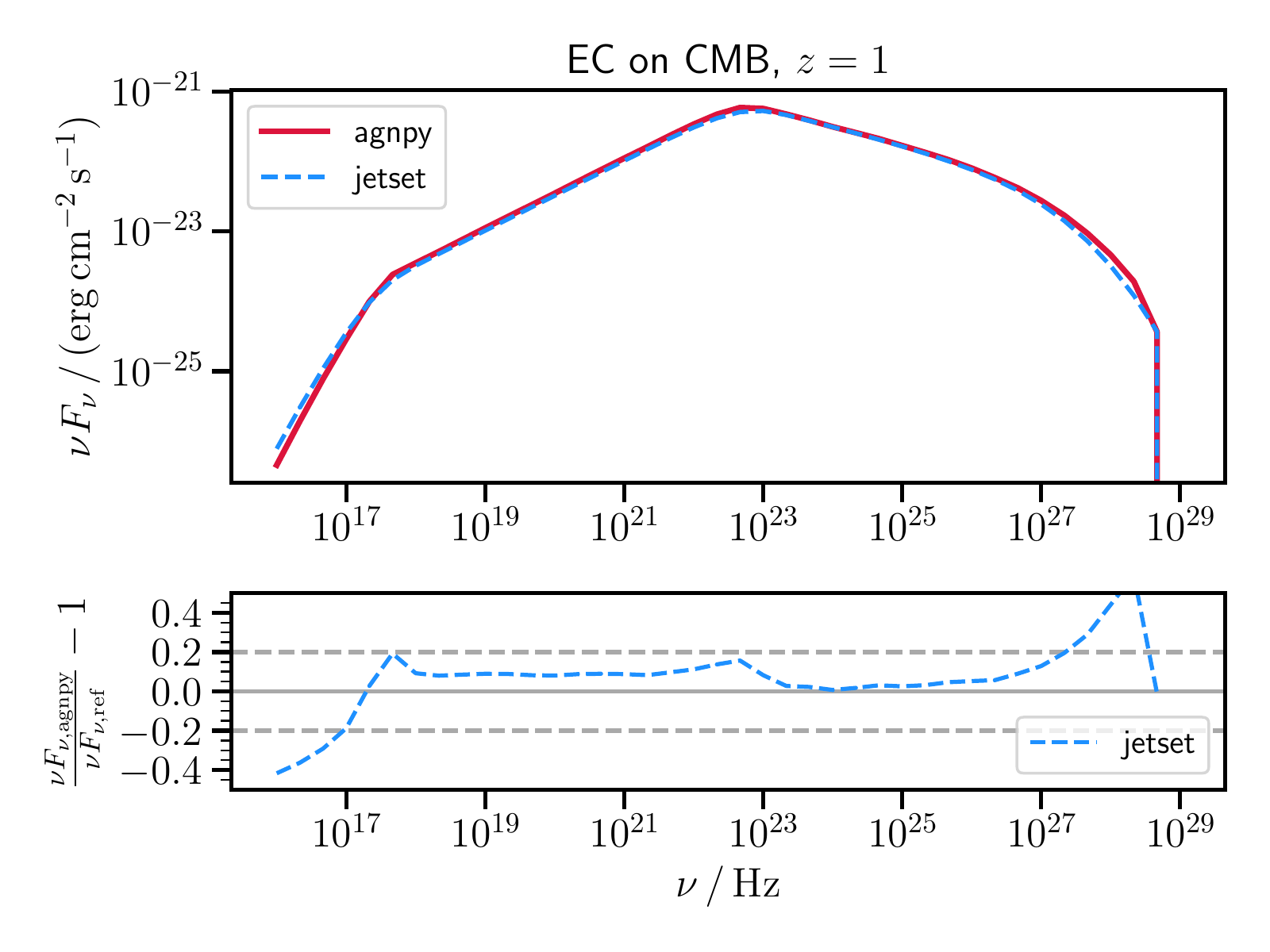}}
\caption{Comparison of the EC scattering on the CMB at $z=1$ computed with \agnpy (red solid line) and jetset (blue dashed line). Fluxes and frequencies are given in the observer reference frame.}
\label{fig:ec_cmb_crosscheck}
\end{figure}
For the EC, we performed comparisons against the literature reproducing the SEDs for the scattering of the accretion disc, BLR, and DT photons in Figs.~8, 10, and 11, respectively, of \citet{finke_2016}. We compared \agnpy against \jetset considering EC on the aforementioned photon fields and additionally on the CMB. For the Compton scattering calculation \jetset relies on the same \cite{gkm_2001} approach used in \agnpy, but the two software packages rely on different parametrisations and assumptions for the target photon fields, therefore larger differences are to be expected in their comparison. As already mentioned, for the EC scenario it is possible to perform an additional internal consistency check approximating, for large distances from the BH, the photon field of the line or thermal emitter with the one produced by a point-like source.
\par
Figures~\ref{fig:ec_disc_cosscheck}-\ref{fig:ec_dt_crosscheck} present, for each AGN photon field, two comparisons. In the first, considering a distance between the blob and the BH smaller than the emitter size, $r < R_{\rm emitter}$, we compare \agnpy against the corresponding bibliographic reference and against \jetset. In the second, considering a significantly larger distance, $r \gg R_{\rm emitter}$, instead of the literature, we compare against the SED computed with \agnpy approximating the emitter with a point-like source. We notice that the point source we consider is monochromatic (Sect.~\ref{subsec:ps_behind_jet}) hence the approximation is exact only in the case in which the emitter itself is also considered monochromatic in the scattering calculation; this is true for the BLR and DT, but not for the disc. The latter has an emission dependent on the radial coordinate $\epsilon_0(R)$ (see Eq.~\ref{eq:epsilon_R}). In considering EC scattering on its radiation (Fig.~\ref{fig:ec_disc_cosscheck}, right panel) we propose a comparison against two point sources: the first emitting with the same energy emitted at the inner disc radius $\epsilon_0 = \epsilon_0(R_{\rm in})$ and the second emitting with the same energy emitted at the outer disc radius $\epsilon_0 = \epsilon_0(R_{\rm out})$. Figure~\ref{fig:ec_cmb_crosscheck} shows instead the comparison against \jetset for EC on the CMB.
\par
For the EC on disc we obtain a deviation within $30\%$, both from the literature and the point-source approximation, for most of the frequency range considered. The accretion disc radiation in \jetset is modelled following \cite{frank_2002}, considering a multi-temperature black body with a temperature profile similar to the one of \citet{dermer_2002}. Deviation from jetset are larger over the entire energy range, with the peaks of the spectra deviating more than $50\%$. For the EC on BLR again we observe a deviation within $30\%$ from the literature and even smaller from the point-source approximation. The BLR radiation in \jetset is modelled, following \cite{donea_2003}, with a thick spherical shell re-emitting the BB radiation of the disc. The different BLR models result in a factor 2 difference between \agnpy and \jetset, over most of the energy range considered. For large distances the deviation is reduced to within $50\%$. Similarly to the previous cases, for the EC on DT we observe  a very small deviation, mostly within $20\%$, from the literature and point-source approximation and a larger one, within $50\%$, from \jetset. The latter accounts in the EC scattering for the whole BB emission of the DT (approximated instead as monochromatic in the \agnpy calculation). For the EC on CMB, no literature reference was available and the two software agree within $30\%$.
\par
For all the EC validations, the parameters of the blob are given under the EC model column of Table~\ref{tab:blob_parameters} and those of the targets in Table~\ref{tab:target_parameters}. The distances $r$ are specified in the plot titles.

\subsection{$\gamma\gamma$ absorption}

\begin{figure*}
\centering
    \includegraphics[width=17cm]{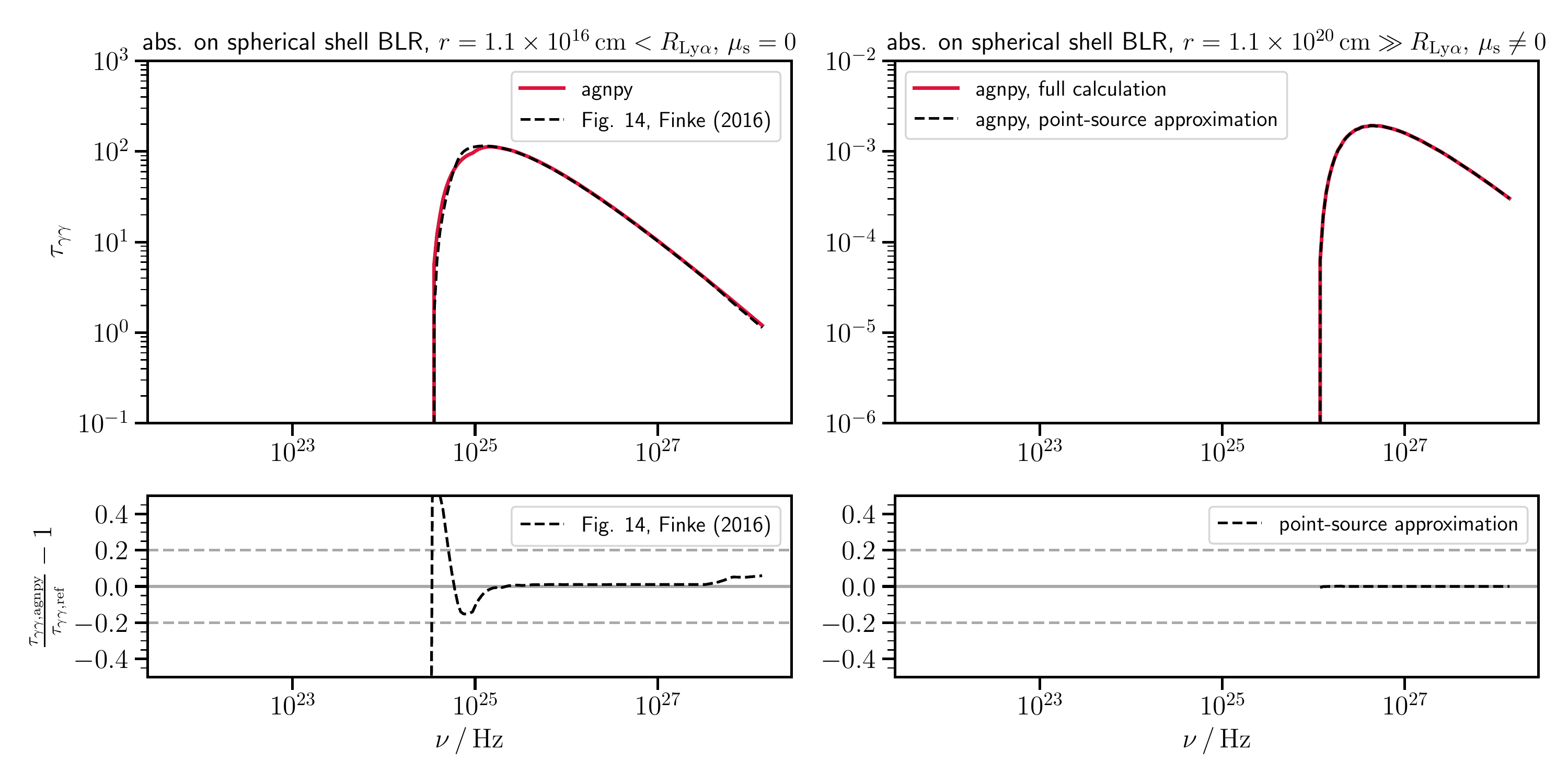}
    \caption{Left: comparison, for a distance smaller than the emitter size, of the $\gamma\gamma$ absorption on the BLR photon field computed by \agnpy (red solid line) against the results in \citet{finke_2016} (black dashed line). Right: comparison, for a distance much larger than the emitter size, of the opacity obtained with the full calculation using the BLR energy density (solid red line) against the calculation approximating the target with a monochromatic point source (black dashed line). To make the opacity for the point-source non null a viewing angle of $\theta_{\rm s} = 20^{\circ}$ is assumed. The bottom panels display the deviation of \agnpy from the reference (or the approximation). Frequencies are given in the observer reference frame.}
    \label{fig:tau_blr_crosscheck}
\end{figure*}

\begin{figure*}
\centering
    \includegraphics[width=17cm]{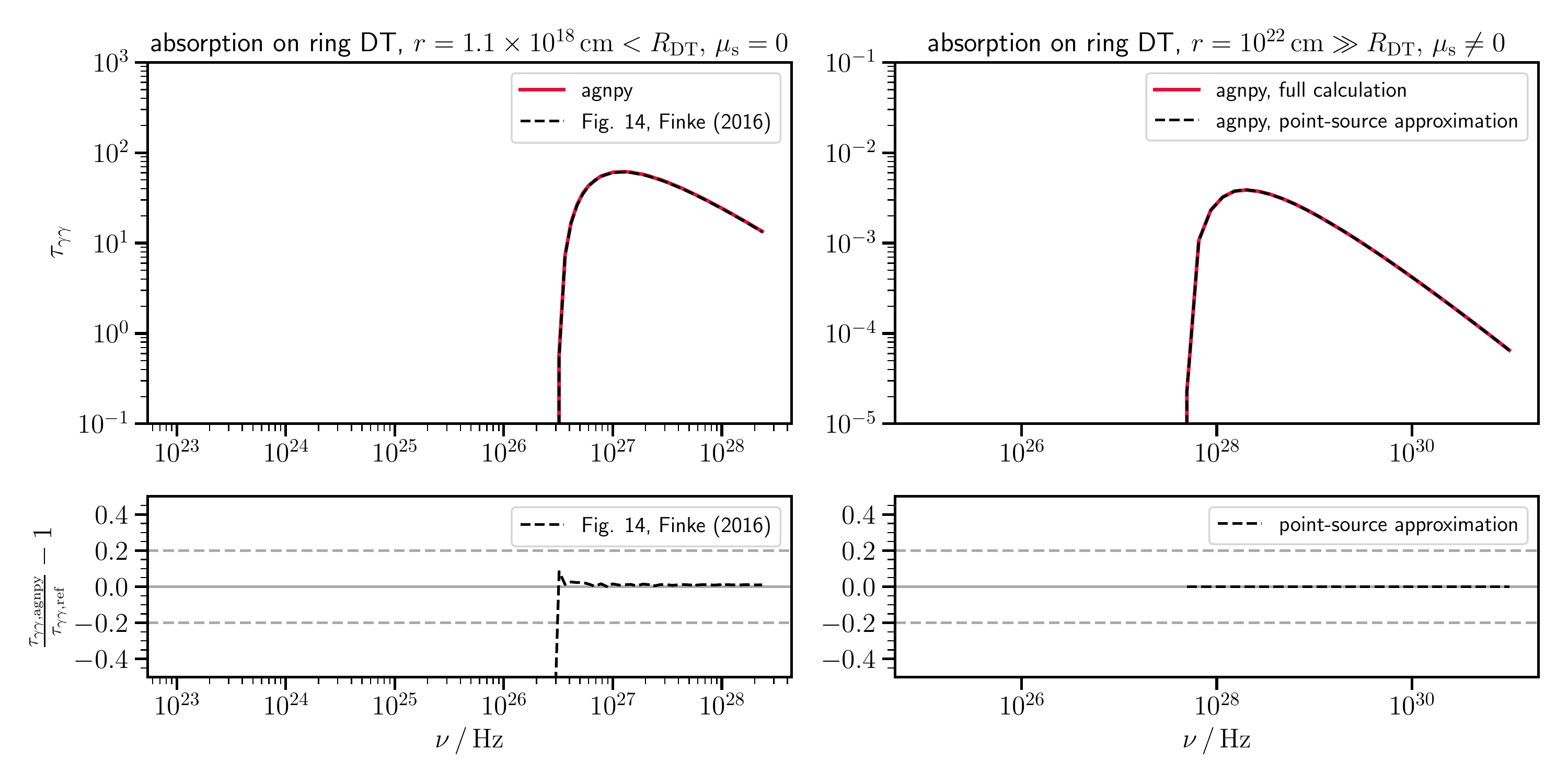}
    \caption{Same as in Fig.~\ref{fig:tau_blr_crosscheck}, but considering $\gamma\gamma$ absorption on the DT photon field.}
    \label{fig:tau_dt_crosscheck}
\end{figure*}

Similar tests performed for the EC were repeated to validate the $\gamma\gamma$ absorption produced by the BLR and DT photon fields, though in this case no comparison with other software was possible. Again, for the distances smaller than the size of the line or thermal emitter, $r < R_{\rm emitter}$, we compared against the corresponding bibliographic reference, that is Fig.~14 of \citet{finke_2016}. For large distances, $r \gg R_{\rm emitter}$, we performed an internal comparison, approximating the target with a point source. We notice that, in the case of a source aligned with the observer, the absorption from a point-source behind the jet would be zero (see Eq.~\ref{eq:tau_ps_behind_jet}). Therefore to perform a meaningful comparison for the absorption due to a point-source behind the jet approximating the emitter, we assumed a viewing angle of $\theta_{\rm s} = 20^{\circ}$ in the case of large distances. Again, with the exception of the extremes of the frequency range considered, we achieve a deviation smaller than $10\%$ in all cases. We did not encounter any bibliographic or software reference to validate the absorption on synchrotron photons.

\section{Reproducibility and performance}
\label{sec:reproducibility}

\begin{table}
\begin{minipage}{\linewidth}
\caption{Execution times of the different computations performed by \agnpy in the validation scripts associated with this repository. \protect\footnote{$n_{\nu}$ represents the number of frequency points over which we compute the SED (or absorption), $n_{\gamma}$ the number of points considered for the EED energies (for SEDs computations), $n_{\theta}$ and $n_{\phi}$ the number of points in zenith and azimuth, respectively, considered for spatial integration. Finally, in the case of absorption, the number of points in distance $n_{l}$ considered for the opacity calculation are indicated.}}
\label{tab:exec_times}      

\begin{tabular}{lrrrrrr}          
\hline\hline
computation & \multicolumn{5}{c}{integration grid} & $t_{\rm exec} / s$ \\
            & $n_{\nu}$ & $n_{\gamma}$ & $n_{\theta}$ & $n_{\phi}$ & $n_{l}$ & \\
\hline
synchrotron & $100$ & $200$ & -- & -- & -- & $0.007$ \\ 
SSC & $100$ & $200$ & -- & -- & -- & $0.404$ \\ 
EC on disc & $40$ & $400$ & $100$ & $50$ & -- & $8.738$ \\
EC on BLR & $40$ & $600$ & $100$ & $50$ & -- & $13.255$ \\
EC on DT & $40$ & $600$ & -- & $50$ & -- & $0.161$ \\
EC on point source & $40$ & $10^3$ & -- & -- & -- & $0.007$\\
EC on CMB & $40$ & $300$ & $100$ & $50$ & -- & $4.984$\\
abs. on BLR & $199$ & -- & $100$ & $50$ & $50$ & $6.044$ \\
abs. on DT & $99$ & -- & -- & $50$ & $50$ & $0.023$ \\
abs. on point source & $199$ & -- & -- & -- & $50$ & $0.013$ \\
\hline
\hline
\end{tabular}
\end{minipage}
\end{table}

The issue of reproducibility in science is often addressed from the point of view of data availability, reduction, and analysis \citep[e.g.][]{chen_2018}, but never from the point of view of physical interpretation of results. Nonetheless, since the latter is dependent on increasingly complex software, it is appropriate to apply the same standards we require for modern data-analysis software to modelling tools. For this reason, beside exposing the source code and opening the development of our software, we took the effort of thoroughly validating its results against bibliographic references, other software and internal consistency checks. Our package offers a modelling tool easily interfaceable with other python-based packages for astrophysical data analysis (\gammapy, \sherpa), thus providing a modular approach to perform a complete and reproducible astrophysical analysis, from data reduction to physical interpretation, in the same computational environment. In order to also make the results in this publication reproducible, we made available all the scripts used to produce them in a \texttt{GitHub} repository\footnote{\href{https://github.com/cosimoNigro/agnpy_paper}{https://github.com/cosimoNigro/agnpy\_paper}}, that we also archived on Zenodo \citep{agnpy_paper_zenodo}. We provide an \texttt{anaconda} environment freezing the version of the package (and dependencies) use and a \texttt{Docker} container\footnote{\href{https://hub.docker.com/r/cosimonigro/agnpy_paper}{https://hub.docker.com/r/cosimonigro/agnpy\_paper}} archiving the computational environment used to produce all the results in this paper.
\par
To provide the reader an estimate of \agnpy's performance in its current state, each script associated with this publication times the most relevant computations performed with the package functions. In Table~\ref{tab:exec_times} we report the execution times, $t_{\rm exec}$, for the radiative processes computations in the validation scripts (those producing Figs.~\ref{fig:ssc_crosscheck}-\ref{fig:tau_dt_crosscheck}). They are obtained the on a \texttt{macOS} system (\texttt{Monterey 12.0.1}) with a $1.4\,{\rm GHz}$ processor and $8\,{\rm GB}$ RAM. For the calculation of a given processes, $t_{\rm exec}$ will depend on the number of dimensions of the integral and on the number of points chosen along each axis when the latter is discretised. We indicate in the table the number of points chosen over each dimension when computing the SED (or the absorption). The number along each dimension is chosen to obtain smooth spectra; we provide further discussion on the effects of the resolution of the discretisation on the final spectra in App.~\ref{sec:discretisation}. The fastest computations are given by one-dimensional integrals, $t_{\rm exec} \approx 10\,{\rm ms}$, such as those estimating the spectra due to synchrotron radiation (Eq.~\ref{eq:synch_sed}) and EC on point-like source (Eq.~\ref{eq:ec_ps_behind_jet_sed}). Two-dimensional integrals like those computing the SSC (Eq.~\ref{eq:ssc_sed}) and EC on DT spectra (Eq.~\ref{eq:ec_dt_sed}) require up to $100\,{\rm ms}$. Finally, the largest three-dimensional integrals such as the spectra of EC on disc (Eq.~\ref{eq:ec_blr_sed}), EC on BLR (Eq.~\ref{eq:ec_ss_disc_sed}), EC on CMB (Eq.~\ref{eq:ec_iso_sed}) can take up to few ${\rm s}$. We remark that the execution times are measured on a single run and might significantly differ from those reported for different machines. They should not differ significantly, mostly at the ${\rm ms}$ level, from execution to execution, on the same machine. The $\chi^2$ minimisation for the Mrk~421 case ($6$ model parameters, $86$ flux points) requires $\approx 15\,{\rm s}$, when using \sherpa and $\approx 120\,{\rm s}$ when using \gammapy. The $\chi^2$ minimisation for the PKS~1510-089 case ($5$ model parameters, $41$ flux points) requires $\approx 23\,{\rm s}$ when using \sherpa and $\approx 85\,{\rm s}$ when using \gammapy.

\section{Conclusions}
\label{sec:conclusions}
In this article we presented \agnpy: a python package modelling the radiative processes in jetted active galaxies. Built on \numpy, \scipy, and \astropy, \agnpy provides another element to the modular astrophysical software system envisioned in \citet{portgies_2018}. Not only it can be easily wrapped by other astrophysical libraries for data analysis living in the same ecosystem, such as \sherpa and \gammapy, but it also complements the science cases covered by other modelling tools such as \naima and \jetset. 
\par
We present several examples of applications, the most useful being the calculation of the spectrum produced by a given radiative process and the computation of the energy densities and the $\gamma\gamma$ absorption on the photon fields of several AGN components. The features connected with line and thermal emitters and with absorption are not included in any other jetted AGN modelling software, nor is the possibility to describe any jetted radio galaxy (including non blazars). Even though not capable, in its current state, of properly describing the time-evolution of the EED, \agnpy provides a class constraining its energies based on a simple parametrisation of the acceleration, escape, and radiation processes. In the spirit of the interfaceability previously discussed, we illustrate, in the code associated with this publication and in the online documentation, how to wrap \agnpy with \sherpa and \gammapy in two examples, fitting the MWL SED of the blazars Mrk~421 and PKS~1510-089.
\par
We provide an extensively and continuously validated modelling tool. Extensively because its results have been validated against bibliographic references, other modelling software and through internal consistency checks. Continuously because the same checks illustrated in Sect.~\ref{sec:crosschecks} constitute only a part of the tests executed during the CI cycle. For all the radiative processes considered, we obtain results that deviate between $10$ and $30\%$ from the bibliographic references and \jetset, when considering the same physical assumptions. Larger differences, still within a factor $0.5-2$, are observed when comparing with the calculations of \jetset that assume a different parametrisations of the photon fields. To our knowledge, \agnpy is the first open-source modelling package for radio-to-gamma-ray non-thermal emission openly providing software and bibliographic comparisons. It would be interesting, provided the availability of other authors, to expand them to results from other publications. Benchmark models for non-thermal radiative processes could be created, as already successfully done, for example, for radiative transfer of photons in dust structures \citep{ivezic_1997, gordon_2017}. Concerning performance, the most simple computations can be executed on current machines in few ${\rm ms}$, while the most complex can take up to few ${\rm s}$. Current performance results in a reasonable execution time, of the order of few seconds, for fitting routines wrapping the code.
\par
Concerning future expansion of the package, we outline two possible paths. The first would be, given the recent multi-messenger observations of blazars \citep{icecube_2018}, and the growing interest in hadronic radiative processes, their implementation in \agnpy. The second might be the addition of a procedure to correctly evolve the EED properly accounting for electrons acceleration, radiation and escape \citep{kardashev_1962, chiaberge_1999}.
\par
With the advent of the next generation of high-energy observatories and the amount of data they will facilitate, open-source modelling tools will become essential to extending the interpretation effort to the largest possible number of astrophysicists. Preparing for widespread access to these data, high-energy astrophysicists have started to develop standardised data formats and open-source analysis tools. In the next era of computational astrophysics the same standardisation and reproducibility standards should be required from modelling software.

\section*{Acknowledgements}
We are grateful to Charles Dermer and Justin Finke for patiently addressing our questions and for providing the spectral points from their publications. We are indebted with Justin Finke and Matteo Cerruti for reviewing this manuscript. We thank Jose Enrique Ruiz for providing comments on an earlier version of the paper and for helping to set up the reproducibility assets. We thank the journal referee for carefully examining the manuscript and the software and for helping us improve the paper.
\par
This work was supported by the European Commission’s Horizon 2020 Program under grant agreement 824064 (ESCAPE - European Science Cluster of Astronomy \& Particle Physics ESFRI Research Infrastructures), by the the ERDF under the Spanish Ministerio de Ciencia e Innovaci{\'o}n (MICINN, grant PID2019-107847RB-C41), and from the CERCA program of the Generalitat de Catalunya.
\par
JS and PG are supported by Polish Narodowe Centrum Nauki grant 2019/34/E/ST9/00224.

\bibliographystyle{aa} 
\bibliography{biblio.bib} 

\begin{appendix}

\section{Notations and formulae}
\label{sec:formulae}
This appendix describes the notation and formulae used in the code implementation and in this paper. As some of these formulae are not explicit in the references given for the specific radiative process implementations (see Section \ref{sec:processes}), we reworked them ourselves. We collect them here hoping they will be useful for cross-checks and future work.
\par 
Following the notation in \citet{dermer_menon_2009} we express the energy of the electrons in terms of their Lorentz factor $\gamma$. For the photons we use a dimensionless energy, $\epsilon$, expressed in units of electron rest-mass energy: $\epsilon = h\nu / (m_{\rm e}c^2)$, where $h$ is the Planck constant, $\nu$ the frequency of the photon, and $m_{\rm e}c^2$ the electron rest-mass energy. Among function arguments, a semicolon separates continuous from parametric dependency, for example the one-dimensional Heaviside function between the points $x=a$ and $x=b$ is written as $H(x; a, b)$. 
\par
The notation we adopt makes frequent use of implicit differentials. We mark them by underlining the function symbol and moving the differential variable among the function arguments, for example the density of electrons with Lorentz factor between $\gamma$ and $\gamma + \diff\gamma$ will be indicated with $\underline{n}_{\rm e}(\gamma) \diff\gamma \equiv \frac{\diff n_{\rm e}}{\diff{\gamma}}\diff\gamma$. Similarly, the specific spectral energy density at a given distance $r$ from a photon field, that is the density of photons with energy between $\epsilon$ and $\epsilon + \diff\epsilon$ and solid angle between $\Omega$ and $\Omega + \diff\Omega$, will be indicated with $\underline{u}(\epsilon, \Omega; r) \diff\epsilon \diff\Omega \equiv \frac{\partial u}{\partial \epsilon  \, \partial \Omega}(r) \diff\epsilon \diff\Omega$. 
\par
The emission region is assumed to be a spherical plasmoid (blob) with radius $R_{\rm b}$, and tangled uniform magnetic field $B$. It is considered homogeneously exposed to the line and thermal emitters radiation (i.e. it is approximated as a point in the EC calculations). The blob moves along the jet axis with (constant) relativistic velocity $\Beta$ and bulk Lorentz factor $\Gamma$, and the jet is oriented at an angle $\theta_{\rm s}$ with respect to the observer line of sight. The Doppler factor, regulating the relativistic transformations to and from the blob's reference frame, is 
\begin{equation}
\delta_{\rm D} = \frac{1}{\Gamma(1-\Beta\cos\theta_s)}.
\label{eq:doppler_factor}
\end{equation}
\par
In this appendix we express physical quantities in three reference frames. Unprimed quantities ($\epsilon$, $\gamma$) are measured in a reference frame comoving with the galaxy and whose origin is the galactic BH. Primed quantities ($\epsilon'$, $\gamma'$) refer instead to a reference frame comoving with the blob. Transformations of energies between these two reference frames read $\epsilon / \epsilon' = \gamma / \gamma' = \delta_{\rm D}$. The cosine of an angle $\mu$, transforms instead as $\mu' = (\mu - \Beta) / (1 - \Beta\mu)$. Hatted quantities refer to the inertial frame of the observer. An observed photon energy, $\hat{\epsilon}$, converts to the galactic reference frame via $\epsilon = \hat{\epsilon}\,(1 + z)$ and to the blob reference frame via $\epsilon' = \hat{\epsilon}\,(1 + z) / \delta_{\rm D}$. We indicate the energy flux (or SED) with the short notation $f_{\epsilon} = \nu F_{\nu}\,[\sedunits]$ where $\epsilon$ is the dimensionless energy corresponding to $\nu$. Observed energy fluxes (i.e. measured at Earth) relate to the intrinsic ones produced in the blob frame via $\hat{f}_{\hat{\epsilon}} = \delta_{\rm D}^4 f'_{\epsilon'}$. All of the plots in the main text of this paper present observed frequencies and fluxes; nonetheless we do not use the hatted notation not to overburden the notation appearing in them. We employ the proper notation in this appendix to mark the transformations between the different reference frames.

\subsection{Non-thermal electron spectra}
\label{sec:electron_densities}
The electron energy density (EED) in the blob reference frame, $\edensity\,[{\rm cm}^{-3}]$, is parametrised as a function of the Lorentz factor only, $\gamma'$, thus assuming electrons with a uniform and isotropic distribution in the emission region. The total volume density of electrons is $n'_{\rm e, \,tot} = \int_{1}^{\infty}\diff\gamma'\,\edensity(\gamma')$. The differential number of particles, $\enumber(\gamma')$, can be obtained multiplying the particle density, $\underline{n}'_{\rm e}(\gamma')$, by the volume of the emission region, $V'_{\rm b}$. The total energy in electrons, $W'_{\rm e}$, is thus defined as
\begin{equation}
W'_{\rm e} = m_{\rm e}c^2 \, \int_{1}^{\infty}\diff\gamma'\,\gamma'\,\enumber(\gamma').
\label{eq:W_e}
\end{equation}
The simplest parametrisation available in \agnpy for the electron energy density is a power law
\begin{equation}
\edensity(\gamma') = k_{\rm e} \, \gamma'^{-p} \, H(\gamma'; \gamma'_{\rm min}, \gamma'_{\rm max}),
\label{eq:power_law}
\end{equation}
where $k_{\rm e}$ is the normalisation constant, $p$ the spectral index and the Heaviside function ensures null values outside the range $\gamma'_{\rm min} \leq \gamma' \leq \gamma'_{\rm max}$. Another parametrisation available is a power law with a spectral index changing from $p_1$ to $p_2$ at the Lorentz factor $\gamma'_{\rm b}$, commonly known as broken power law
\begin{equation}
\begin{split}
\edensity(\gamma') = k_{\rm e} 
&\left[ \left(\frac{\gamma'}{\gamma'_{\rm b}}\right)^{-p_1} H(\gamma'; \gamma'_{\rm min}, \gamma'_{\rm b}) \right.\\   
&\left. + \left(\frac{\gamma'}{\gamma'_{\rm b}}\right)^{-p_2} H(\gamma'; \gamma'_{\rm b}, \gamma'_{\rm max}) \right].\\
\end{split}
\label{eq:broken_power_law}
\end{equation}
A power law with an exponential cutoff at the Lorentz factor $\gamma'_{\rm c}$ is also available
\begin{equation}
\edensity(\gamma') = k_{\rm e} \, \gamma'^{-p} \, \exp\left(-\frac{\gamma'}{\gamma'_{\rm c}}\right) \, H(\gamma'; \gamma'_{\rm min}, \gamma'_{\rm max}). 
\label{eq:exp_cutoff_power_law}
\end{equation}
Finally, a curved power law, with spectral index changing logarithmically with the Lorentz factor, commonly known as log-parabola, can also be considered
\begin{equation}
\edensity(\gamma') = k_{\rm e} \, \left(\frac{\gamma'}{\gamma'_0}\right)^{-(p + q \log_{10}(\gamma' / \gamma'_0))} \, H(\gamma'; \gamma'_{\rm min}, \gamma'_{\rm max}) . 
\label{eq:log_parabola}
\end{equation}

\subsection{Energy densities of line and thermal emitters}
\label{sec:energy_densities}

\begin{figure}
\centering
  
\tikzset {_qjpqod523/.code = {\pgfsetadditionalshadetransform{ \pgftransformshift{\pgfpoint{0 bp } { 0 bp }  }  \pgftransformscale{1 }  }}}
\pgfdeclareradialshading{_q9hlx3qwl}{\pgfpoint{0bp}{0bp}}{rgb(0bp)=(1,1,1);
rgb(0bp)=(1,1,1);
rgb(25bp)=(0,0,0);
rgb(400bp)=(0,0,0)}

  
\tikzset {_a0v358y05/.code = {\pgfsetadditionalshadetransform{ \pgftransformshift{\pgfpoint{-198 bp } { -198 bp }  }  \pgftransformscale{1.32 }  }}}
\pgfdeclareradialshading{_8v1h1yirn}{\pgfpoint{160bp}{160bp}}{rgb(0bp)=(1,1,1);
rgb(0bp)=(1,1,1);
rgb(25bp)=(0.48,0.15,0.15);
rgb(400bp)=(0.48,0.15,0.15)}
\tikzset{every picture/.style={line width=0.75pt}} 

\begin{tikzpicture}[x=0.75pt,y=0.75pt,yscale=-1,xscale=1]

\draw    (288.73,257.36) -- (469,257.36) ;
\draw [shift={(471,257.36)}, rotate = 180] [color={rgb, 255:red, 0; green, 0; blue, 0 }  ][line width=0.75]    (10.93,-3.29) .. controls (6.95,-1.4) and (3.31,-0.3) .. (0,0) .. controls (3.31,0.3) and (6.95,1.4) .. (10.93,3.29)   ;
\draw    (288.73,257.36) -- (171.59,325.15) ;
\draw [shift={(169.86,326.15)}, rotate = 329.94] [color={rgb, 255:red, 0; green, 0; blue, 0 }  ][line width=0.75]    (10.93,-3.29) .. controls (6.95,-1.4) and (3.31,-0.3) .. (0,0) .. controls (3.31,0.3) and (6.95,1.4) .. (10.93,3.29)   ;
\draw    (288.73,257.36) -- (288.01,32) ;
\draw [shift={(288,30)}, rotate = 449.82] [color={rgb, 255:red, 0; green, 0; blue, 0 }  ][line width=0.75]    (10.93,-3.29) .. controls (6.95,-1.4) and (3.31,-0.3) .. (0,0) .. controls (3.31,0.3) and (6.95,1.4) .. (10.93,3.29)   ;
\draw   (280.97,98.84) .. controls (276.3,98.83) and (273.96,101.15) .. (273.95,105.82) -- (273.75,166.44) .. controls (273.72,173.11) and (271.38,176.43) .. (266.71,176.42) .. controls (271.38,176.43) and (273.7,179.77) .. (273.68,186.44)(273.69,183.44) -- (273.48,247.07) .. controls (273.47,251.74) and (275.79,254.08) .. (280.46,254.1) ;
\draw  [dash pattern={on 4.5pt off 4.5pt}]  (252.28,44) -- (401.29,257.79) ;
\draw  [line width=1.5]  (168,262.41) .. controls (168,231.8) and (222.51,206.98) .. (289.76,206.98) .. controls (357,206.98) and (411.52,231.8) .. (411.52,262.41) .. controls (411.52,293.02) and (357,317.84) .. (289.76,317.84) .. controls (222.51,317.84) and (168,293.02) .. (168,262.41) -- cycle ;
\draw  [line width=1.5]  (189.88,261.44) .. controls (189.88,237.94) and (235.07,218.89) .. (290.82,218.89) .. controls (346.56,218.89) and (391.75,237.94) .. (391.75,261.44) .. controls (391.75,284.95) and (346.56,304) .. (290.82,304) .. controls (235.07,304) and (189.88,284.95) .. (189.88,261.44) -- cycle ;
\draw    (288.73,147.18) .. controls (301.88,151.34) and (310.93,147.06) .. (318.57,140.47) ;
\draw    (267.23,66.1) .. controls (273.29,59.93) and (275.47,56.14) .. (288.44,55.52) ;
\draw   (289.76,264.26) .. controls (289.76,268.93) and (292.09,271.26) .. (296.76,271.26) -- (335.06,271.26) .. controls (341.73,271.26) and (345.06,273.59) .. (345.06,278.26) .. controls (345.06,273.59) and (348.39,271.26) .. (355.06,271.26)(352.06,271.26) -- (393.36,271.26) .. controls (398.03,271.26) and (400.36,268.93) .. (400.36,264.26) ;
\draw  [dash pattern={on 4.5pt off 4.5pt}]  (288.73,257.36) -- (253.51,328) ;
\draw    (261.87,273.5) .. controls (270,279) and (267,278) .. (277,278) ;
\path  [shading=_q9hlx3qwl,_qjpqod523] (282.23,257.36) .. controls (282.23,253.77) and (285.14,250.86) .. (288.73,250.86) .. controls (292.32,250.86) and (295.23,253.77) .. (295.23,257.36) .. controls (295.23,260.95) and (292.32,263.86) .. (288.73,263.86) .. controls (285.14,263.86) and (282.23,260.95) .. (282.23,257.36) -- cycle ; 
 \draw   (282.23,257.36) .. controls (282.23,253.77) and (285.14,250.86) .. (288.73,250.86) .. controls (292.32,250.86) and (295.23,253.77) .. (295.23,257.36) .. controls (295.23,260.95) and (292.32,263.86) .. (288.73,263.86) .. controls (285.14,263.86) and (282.23,260.95) .. (282.23,257.36) -- cycle ; 

\draw [color={rgb, 255:red, 208; green, 2; blue, 27 }  ,draw opacity=1 ] [dash pattern={on 4.5pt off 4.5pt}]  (288.73,98.36) -- (315.23,34.85) ;
\draw [shift={(316,33)}, rotate = 472.65] [color={rgb, 255:red, 208; green, 2; blue, 27 }  ,draw opacity=1 ][line width=0.75]    (10.93,-3.29) .. controls (6.95,-1.4) and (3.31,-0.3) .. (0,0) .. controls (3.31,0.3) and (6.95,1.4) .. (10.93,3.29)   ;
\path  [shading=_8v1h1yirn,_a0v358y05] (282.23,98.36) .. controls (282.23,94.77) and (285.14,91.86) .. (288.73,91.86) .. controls (292.32,91.86) and (295.23,94.77) .. (295.23,98.36) .. controls (295.23,101.95) and (292.32,104.86) .. (288.73,104.86) .. controls (285.14,104.86) and (282.23,101.95) .. (282.23,98.36) -- cycle ; 
 \draw  [color={rgb, 255:red, 0; green, 0; blue, 0 }  ,draw opacity=1 ] (282.23,98.36) .. controls (282.23,94.77) and (285.14,91.86) .. (288.73,91.86) .. controls (292.32,91.86) and (295.23,94.77) .. (295.23,98.36) .. controls (295.23,101.95) and (292.32,104.86) .. (288.73,104.86) .. controls (285.14,104.86) and (282.23,101.95) .. (282.23,98.36) -- cycle ; 

\draw [color={rgb, 255:red, 208; green, 2; blue, 27 }  ,draw opacity=1 ]   (288.44,55.52) .. controls (294,57) and (299,60) .. (302.37,65.68) ;

\draw (261.92,173.25) node    {$r$};
\draw (273.4,45.35) node    {$\theta $};
\draw (307.51,158.01) node    {$\theta $};
\draw (344.83,285.31) node    {$R$};
\draw (258.78,285.5) node    {$\phi $};
\draw (298.22,46) node  [color={rgb, 255:red, 208; green, 2; blue, 27 }  ,opacity=1 ]  {$\theta_s $};
\draw (342.18,21.5) node   [align=left] {\textcolor[rgb]{0.82,0.01,0.11}{observer}};
\draw (320.18,97.5) node   [align=left] {blob};
\draw (306.42,242.5) node   [align=left] {BH};

\end{tikzpicture}
\caption{Geometry used for the energy density and external Compton scattering calculations. A disc-like emitter with radial coordinate $R$ is represented as an example.}
\label{fig:target_geometry}
\end{figure}
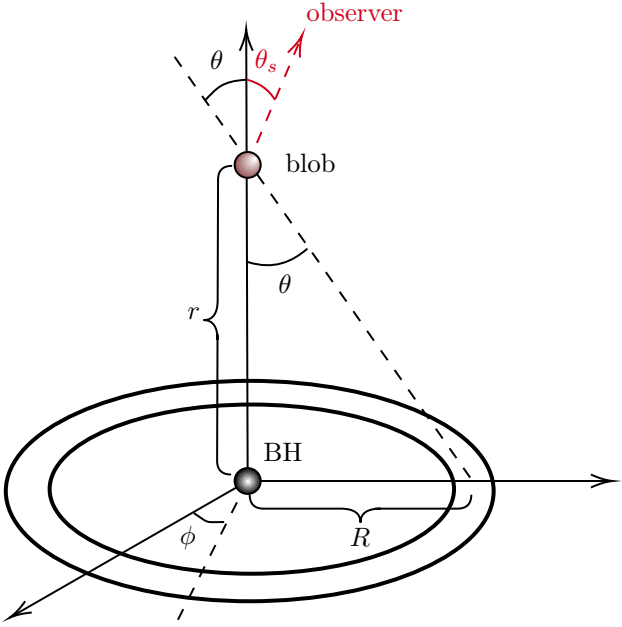

The (integral) energy density of a photon field at a given distance $r$ above the central BH, $u(r)\,[{\rm erg}\,{\rm cm}^{-3}]$, (see Section~\ref{subsec:u_targets}) can be computed by integrating the specific spectral energy density over energy and solid angle
\begin{equation}
u(r) = \int\diff\epsilon \int\diff\Omega \; \phdensity(\epsilon, \Omega; r).
\label{eq:u_r_general}
\end{equation}
Let us consider the reference frame with the BH in its centre and the scattering geometry illustrated in Fig.~\ref{fig:target_geometry}. An illustrative disc-like emitter with variable radius $R$ illuminates the blob, which is streaming along the jet axis and is located at a distance $r$ above the BH. Photons from the emitter hit the blob with zenith angle $\theta$ and azimuth angle $\phi$. The jet axis at the blob position forms a viewing angle $\theta_{\rm s}$ with the line of sight of the observer. In the following section we calculate the energy densities for all the emitters (targets) implemented in \agnpy, both in the galactic ($u$) and in the blob  ($u'$) reference frames. The transformation laws for spectral energy densities between these two reference frames can be found in  \citet[][Section~4]{dermer_2002} and \citet[][Chapter~5]{dermer_menon_2009}.

\subsubsection{Isotropic monochromatic}
An isotropic monochromatic radiation field with dimensionless energy $\epsilon_0$ has specific spectral energy density
\begin{equation}
\phdensity(\epsilon, \Omega) = \frac{u_0\,\delta(\epsilon - \epsilon_0)}{4 \pi},
\label{eq:u_iso}
\end{equation}
such that the energy density in the galactic frame is trivially $u=u_0$, while in the blob frame
\begin{equation}
u' = u_0 \, \Gamma^2 \left(1 + \frac{\Beta^2}{3}\right),   
\end{equation}
in both cases the integral energy density is independent from the distance, as visible in Fig.~\ref{fig:u_targets_example} for the CMB case.

\subsubsection{Monochromatic point source behind the blob}
\label{subsec:ps_behind_jet}
Let us consider a monochromatic ($\epsilon = \epsilon_0$) source of luminosity $L_0$ at the centre of our reference frame (the BH position). Its specific spectral energy density is
\begin{equation}
\phdensity(\epsilon, \Omega; r) = \frac{L_0}{4 \pi c r^2} \frac{\delta(\mu-1)}{2 \pi} \delta(\epsilon - \epsilon_0),
\label{eq:u_ps_behind_jet}
\end{equation}
where the condition $\mu=\cos\theta=1$ ensures that the photons are coming from behind the blob. The integral energy density in the galactic frame is 
\begin{equation}
u(r) = \frac{L_0}{4 \pi c r^2}, 
\label{eq:point_source_stat}
\end{equation}
and in the blob frame
\begin{equation}
u'(r) = \frac{1}{\Gamma^2 (1 + \Beta)^2}\,\frac{L_0}{4 \pi c r^2} = 
        \frac{u(r)}{\Gamma^2 (1 + \Beta)^2}.
\label{eq:point_source_com}
\end{equation}
Where $\Gamma$ and $\Beta$ are the Lorentz factor and velocity (in units of $c$) of the blob. We notice that Eqs.~(\ref{eq:point_source_stat}) and (\ref{eq:point_source_com}) can be used as a cross-check for the correct calculation of the photon energy densities produced by emitters with a more complex geometry. For a large distance from the BH, where any emitter is observed under a very small angle $\theta \ll \theta_{\rm s}$, its photon field tends to the one produced by a point-like source. Hence Eqs.~(\ref{eq:point_source_stat}) and (\ref{eq:point_source_com}) represent limits to the energy densities of any emitter, the limits being exact for monochromatic emitters.

\subsubsection{Shakura-Sunyaev accretion disc}
Let us consider an accretion disc with inner and outer radii $R_{\rm in}$ and $R_{\rm out}$, respectively. From the sketch in Fig.~\ref{fig:target_geometry} we see that the cosine of the photon incoming angle $\mu=\cos\theta$ and the coordinate along the disc radius $R$ are related via
\begin{equation}
\mu = \left(1 + \frac{R^2}{r^2}\right)^{-1/2}, \; 
R = r \sqrt{\mu^{-2} - 1}.
\label{eq:R_from_mu_r}
\end{equation}
The  minimum and maximum incoming angles of photons from the disc impinging on the emission region are therefore
\begin{equation}
\mu_{\rm min} = \left( 1 + \frac{R^2_{\rm out}}{r^2} \right)^{-1/2}, \;
\mu_{\rm max} = \left( 1 + \frac{R^2_{\rm in}}{r^2} \right)^{-1/2}.
\label{eq:mu_limits_ss_disc}
\end{equation}
The specific spectral energy density of photons coming from a \citet{shakura} accretion disc can be parametrised, following \citet{dermer_2002} and \citet{dermer_2009}, as:
\begin{equation}
\phdensity(\epsilon, \Omega; r) = \frac{3 G M \dot{m}}{(4 \pi)^2 c R^3} \, \varphi(R) \, 
                         \delta(\epsilon - \epsilon_0(R))
\label{eq:u_ss_disc}
\end{equation}
where $\varphi(R)$ represents the variation of energy flux along the radius
\begin{equation}
\varphi(R) = 1 - \sqrt{\frac{R_{\rm in}}{R}}
\label{eq:phi_R}
\end{equation}
and $\epsilon_0(R)$ is the monochromatic approximation for the mean photon energy emitted from the disc at radius $R$
\begin{equation}
\epsilon_0(R) = 
    2.7 \times 10^{-4} 
    \left(\frac{l_{\rm Edd}}{M_8 \eta}\right)^{1/4} 
    \left(\frac{R}{R_{\rm g}}\right)^{-3/4}.
\label{eq:epsilon_R}
\end{equation}
In the above equations, $G$ is the gravitational constant, $M = M_8 \times 10^8\,M_{\odot}$ is the mass of the black hole in solar mass units, $\dot{m}\,[{\rm g}\,{\rm s}^{-1}]$ the black hole mass accretion rate, $\eta$ the fraction of gravitational energy converted to radiant energy ($L_{\rm disc} = \eta \dot{m} c^2$), $l_{\rm Edd}$ the ratio of the disc luminosity to the Eddington luminosity ($L_{\rm Edd} = 1.26 \times 10^{46}\,M_8\,{\rm erg}\,{\rm s}^{-1}$), and $R_{\rm g} = GM/c^2$ is the gravitational radius of the BH. When using Eq.~(\ref{eq:phi_R}) and Eq.~(\ref{eq:epsilon_R}), we make explicit the fact that the photons emitted at a given disc radius $R$ will have different incidence angles depending on the blob position $r$, by replacing $\varphi(R) \rightarrow \varphi(\mu; r)$ and $\epsilon_0(R) \rightarrow \epsilon_0(\mu; r)$. \par
The integral energy density in the galactic frame is
\begin{equation}
u(r) = \frac{3}{8 \pi c} \frac{G M \dot{m}}{r^3} 
       \int_{\mu_{\rm min}}^{\mu_{\rm max}}\diff\mu \,
       \frac{\varphi(\mu; r)}{(\mu^{-2} - 1)^{3/2}}.
\label{eq:ssdisc_stat}
\end{equation}
In the blob frame the energy density is
\begin{equation}
\begin{split}
u'(r) =& \frac{3}{8 \pi c} \frac{G M \dot{m}}{r^3}\\
    &\int_{\mu_{\rm min}}^{\mu_{\rm max}}\diff\mu
    \frac{\varphi(\mu; r)}{\Gamma^6 (1 - \Beta\mu)^2 (1 + \Beta \mu')^4 (\mu^{-2} - 1)^{-3/2}}.\\
\end{split}
\label{eq:ssdisc_com}
\end{equation}

\subsubsection{Spherical shell broad line region}

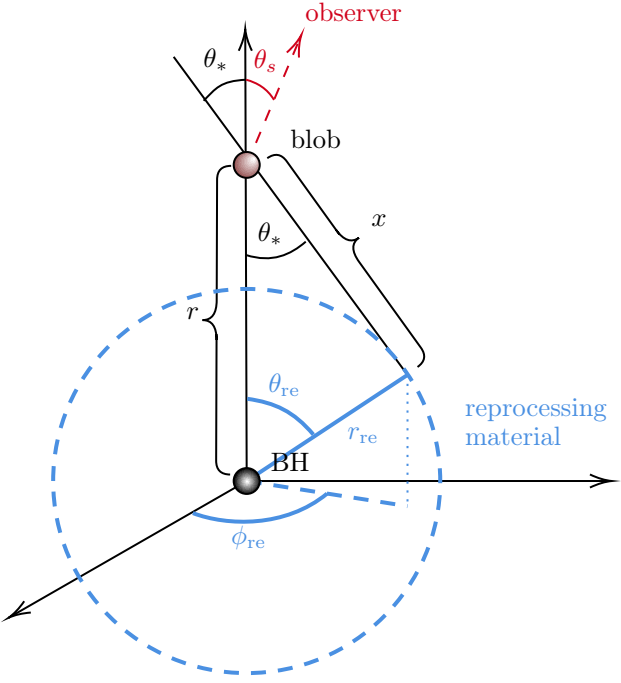
\begin{figure}
\centering
  
\tikzset {_a37f7ugzc/.code = {\pgfsetadditionalshadetransform{ \pgftransformshift{\pgfpoint{-198 bp } { -198 bp }  }  \pgftransformscale{1.32 }  }}}
\pgfdeclareradialshading{_5ae5m99mj}{\pgfpoint{160bp}{160bp}}{rgb(0bp)=(1,1,1);
rgb(0bp)=(1,1,1);
rgb(25bp)=(0.48,0.15,0.15);
rgb(400bp)=(0.48,0.15,0.15)}

  
\tikzset {_etu95k2gm/.code = {\pgfsetadditionalshadetransform{ \pgftransformshift{\pgfpoint{0 bp } { 0 bp }  }  \pgftransformscale{1 }  }}}
\pgfdeclareradialshading{_odltk1kjx}{\pgfpoint{0bp}{0bp}}{rgb(0bp)=(1,1,1);
rgb(0bp)=(1,1,1);
rgb(25bp)=(0,0,0);
rgb(400bp)=(0,0,0)}
\tikzset{every picture/.style={line width=0.75pt}} 

\begin{tikzpicture}[x=0.75pt,y=0.75pt,yscale=-1,xscale=1]

\draw    (288.73,257.36) -- (469,257.36) ;
\draw [shift={(471,257.36)}, rotate = 180] [color={rgb, 255:red, 0; green, 0; blue, 0 }  ][line width=0.75]    (10.93,-3.29) .. controls (6.95,-1.4) and (3.31,-0.3) .. (0,0) .. controls (3.31,0.3) and (6.95,1.4) .. (10.93,3.29)   ;
\draw    (288.73,257.36) -- (171.59,325.15) ;
\draw [shift={(169.86,326.15)}, rotate = 329.94] [color={rgb, 255:red, 0; green, 0; blue, 0 }  ][line width=0.75]    (10.93,-3.29) .. controls (6.95,-1.4) and (3.31,-0.3) .. (0,0) .. controls (3.31,0.3) and (6.95,1.4) .. (10.93,3.29)   ;
\draw    (288.73,257.36) -- (288.01,32) ;
\draw [shift={(288,30)}, rotate = 449.82] [color={rgb, 255:red, 0; green, 0; blue, 0 }  ][line width=0.75]    (10.93,-3.29) .. controls (6.95,-1.4) and (3.31,-0.3) .. (0,0) .. controls (3.31,0.3) and (6.95,1.4) .. (10.93,3.29)   ;
\draw   (280.97,98.84) .. controls (276.3,98.83) and (273.96,101.15) .. (273.95,105.82) -- (273.75,166.44) .. controls (273.72,173.11) and (271.38,176.43) .. (266.71,176.42) .. controls (271.38,176.43) and (273.7,179.77) .. (273.68,186.44)(273.69,183.44) -- (273.48,247.07) .. controls (273.47,251.74) and (275.79,254.08) .. (280.46,254.1) ;
\draw    (252.28,44) -- (369,204) ;
\draw    (288.37,143.68) .. controls (301.52,147.84) and (310.57,143.56) .. (318.21,136.97) ;
\draw    (267.23,66.1) .. controls (273.29,59.93) and (275.47,56.14) .. (288.44,55.52) ;
\draw [color={rgb, 255:red, 74; green, 144; blue, 226 }  ,draw opacity=1 ][line width=1.5]  [dash pattern={on 5.63pt off 4.5pt}]  (288.73,257.36) -- (369,270) ;
\draw [color={rgb, 255:red, 74; green, 144; blue, 226 }  ,draw opacity=1 ][line width=1.5]    (261.87,273.5) .. controls (279,280) and (308,281) .. (328.87,263.68) ;
\draw [color={rgb, 255:red, 208; green, 2; blue, 27 }  ,draw opacity=1 ] [dash pattern={on 4.5pt off 4.5pt}]  (288.73,98.36) -- (315.23,34.85) ;
\draw [shift={(316,33)}, rotate = 472.65] [color={rgb, 255:red, 208; green, 2; blue, 27 }  ,draw opacity=1 ][line width=0.75]    (10.93,-3.29) .. controls (6.95,-1.4) and (3.31,-0.3) .. (0,0) .. controls (3.31,0.3) and (6.95,1.4) .. (10.93,3.29)   ;
\path  [shading=_5ae5m99mj,_a37f7ugzc] (282.23,98.36) .. controls (282.23,94.77) and (285.14,91.86) .. (288.73,91.86) .. controls (292.32,91.86) and (295.23,94.77) .. (295.23,98.36) .. controls (295.23,101.95) and (292.32,104.86) .. (288.73,104.86) .. controls (285.14,104.86) and (282.23,101.95) .. (282.23,98.36) -- cycle ; 
 \draw  [color={rgb, 255:red, 0; green, 0; blue, 0 }  ,draw opacity=1 ] (282.23,98.36) .. controls (282.23,94.77) and (285.14,91.86) .. (288.73,91.86) .. controls (292.32,91.86) and (295.23,94.77) .. (295.23,98.36) .. controls (295.23,101.95) and (292.32,104.86) .. (288.73,104.86) .. controls (285.14,104.86) and (282.23,101.95) .. (282.23,98.36) -- cycle ; 

\draw [color={rgb, 255:red, 208; green, 2; blue, 27 }  ,draw opacity=1 ]   (288.44,55.52) .. controls (294,57) and (299,60) .. (302.37,65.68) ;
\draw  [color={rgb, 255:red, 74; green, 144; blue, 226 }  ,draw opacity=1 ][dash pattern={on 5.63pt off 4.5pt}][line width=1.5]  (192.12,257.36) .. controls (192.12,204) and (235.37,160.75) .. (288.73,160.75) .. controls (342.09,160.75) and (385.34,204) .. (385.34,257.36) .. controls (385.34,310.72) and (342.09,353.98) .. (288.73,353.98) .. controls (235.37,353.98) and (192.12,310.72) .. (192.12,257.36) -- cycle ;
\draw [color={rgb, 255:red, 74; green, 144; blue, 226 }  ,draw opacity=1 ][line width=1.5]    (288.73,257.36) -- (369,204) ;
\draw [color={rgb, 255:red, 74; green, 144; blue, 226 }  ,draw opacity=1 ] [dash pattern={on 0.84pt off 2.51pt}]  (369,204) -- (369,270) ;
\draw [color={rgb, 255:red, 74; green, 144; blue, 226 }  ,draw opacity=1 ][line width=1.5]    (289,216) .. controls (303,218) and (312.13,222.32) .. (322,234) ;
\draw   (374,200) .. controls (377.8,197.29) and (378.34,194.03) .. (375.63,190.24) -- (344.88,147.2) .. controls (341.01,141.77) and (340.97,137.7) .. (344.77,134.99) .. controls (340.97,137.7) and (337.13,136.35) .. (333.26,130.92)(335,133.36) -- (308.76,96.63) .. controls (306.05,92.83) and (302.79,92.29) .. (299,95) ;
\path  [shading=_odltk1kjx,_etu95k2gm] (282.23,257.36) .. controls (282.23,253.77) and (285.14,250.86) .. (288.73,250.86) .. controls (292.32,250.86) and (295.23,253.77) .. (295.23,257.36) .. controls (295.23,260.95) and (292.32,263.86) .. (288.73,263.86) .. controls (285.14,263.86) and (282.23,260.95) .. (282.23,257.36) -- cycle ; 
 \draw   (282.23,257.36) .. controls (282.23,253.77) and (285.14,250.86) .. (288.73,250.86) .. controls (292.32,250.86) and (295.23,253.77) .. (295.23,257.36) .. controls (295.23,260.95) and (292.32,263.86) .. (288.73,263.86) .. controls (285.14,263.86) and (282.23,260.95) .. (282.23,257.36) -- cycle ; 

\draw (261.92,173.25) node    {$r$};
\draw (273.4,45.35) node    {$\theta_\ast $};
\draw (300.51,133.01) node    {$\theta_\ast $};
\draw (346.83,234.31) node  [color={rgb, 255:red, 74; green, 144; blue, 226 }  ,opacity=1 ]  {$r_{\rm re}$};
\draw (307.51,209.01) node  [color={rgb, 255:red, 74; green, 144; blue, 226 }  ,opacity=1 ]  {$\theta_{\rm re}$};
\draw (289.78,285.5) node  [color={rgb, 255:red, 74; green, 144; blue, 226 }  ,opacity=1 ]  {$\phi_{\rm re}$};
\draw (298.22,46) node  [color={rgb, 255:red, 208; green, 2; blue, 27 }  ,opacity=1 ]  {$\theta_s$};
\draw (342.18,21.5) node   [align=left] {\textcolor[rgb]{0.82,0.01,0.11}{observer}};
\draw (323.18,85.5) node   [align=left] {blob};
\draw (310.42,247.5) node   [align=left] {BH};

\draw (354.92,126.25) node    {$x$};
\draw (433.18,227.5) node  [color={rgb, 255:red, 74; green, 144; blue, 226 }  ,opacity=1 ] [align=left] {reprocessing\\material};

\end{tikzpicture}
\caption{Geometry used for the energy density and external Compton scattering calculations in case of a spherically symmetric emitter reprocessing the accretion disc radiation.}
\label{fig:reprocessing_target_geometry}
\end{figure}

Following \citet{finke_2016}, we model the BLR as an infinitesimally thin, monochromatic shell. A complex model of a BLR emitting several lines can be built as a superposition of these simpler objects. Fig.~\ref{fig:reprocessing_target_geometry} presents the geometry of a spherically symmetric emitter reprocessing the accretion disc radiation. The cosine of the incidence angle of photons produced on the surface of the reprocessing material and crossing the jet at a distance $r$ from the BH is $\mu_\ast = \cos\theta_\ast$. The spherical coordinates $(r_{\rm re}, \mu_{\rm re}, \phi_{\rm re})$ represent the reprocessing material (with $\mu_{\rm re} = \cos\theta_{\rm re}$). From geometry
\begin{equation}
\begin{split}
\mu_\ast^2 &= 1 - \left( \frac{r_{\rm re}}{x} \right)^2 (1 - \mu_{\rm re}^2), \\
       x^2 &= r_{\rm re}^2 + r^2 - 2 r  r_{\rm re} \mu_{\rm re}.
\end{split}
\label{eq:reprocessed_geom}
\end{equation}
The specific spectral energy density of a spherical shell BLR reads \citep{finke_2016}
\begin{equation}
\begin{split}
\phdensity(\epsilon, \Omega; r) = 
\frac{\xi_{\rm line} L_{\rm disc}}{(4\pi)^2c} \delta(\epsilon - \epsilon_{\rm line}) 
\int_{-1}^{1}\frac{\diff\mu_{\rm re}}{x^2} \delta(\mu - \mu_\ast),
\end{split}
\label{eq:u_blr}
\end{equation}
where $\xi_{\rm line}$ is the fraction of the disc luminosity ($L_{\rm disc}$) reprocessed by the BLR, $\epsilon_{\rm line}$ is the dimensionless energy of the emitted spectral line, $\mu_\ast$ and $\mu_{\rm re}$ are obtained from Eq.~(\ref{eq:reprocessed_geom}) with $r_{\rm re} = R_{\rm line}$, the radius at which we assume the monochromatic line is emitted. The integral energy density in the galactic frame is
\begin{equation}
u(r) = \frac{\xi_{\rm line} L_{\rm disc}}{8 \pi c} \int_{-1}^{1}\frac{\diff\mu_{\rm re}}{x^2}.
\label{eq:blr_stat}
\end{equation}
The monochromatic spherical shell BLR can be approximated, at a large distance from the BH, as a point source behind the blob, so Eq.~(\ref{eq:blr_stat}) should reduce to Eq.~(\ref{eq:point_source_stat}) in the limit $r \gg R_{\rm line}$. As $x^2 \rightarrow r^2$
\begin{equation}
u(r) = \frac{\xi_{\rm line} L_{\rm disc}}{8 \pi c} \int_{-1}^{1}\frac{\diff\mu_{\rm re}}{r^2} 
     = \frac{\xi_{\rm line} L_{\rm disc}}{4 \pi c r^2},
\end{equation}
which is Eq.~(\ref{eq:point_source_stat}), the energy density of a monochromatic 
point source behind the blob, with luminosity $L_0 = \xi_{\rm line} L_{\rm disc}$.
The integrated energy density in the blob frame reads
\begin{equation}
u'(r) = \frac{\xi_{\rm line} L_{\rm disc}}{8 \pi c}
             \Gamma^2 \int_{-1}^{1} (1 - \Beta\mu_*)^2
             \frac{\diff\mu_{\rm re}}{x^2}.
\label{eq:blr_com}
\end{equation}
We can check again that  in the limit of large distance the latter reduces to the equivalent expression for the point-source behind the blob, Eq.~(\ref{eq:point_source_com}). 
For $r \gg R_{\rm line}$, $x^2 \rightarrow r^2$ and $\mu_{\ast} \rightarrow 1$, therefore
\begin{equation}
u'(r) = \Gamma^2 (1 - \Beta)^2 \frac{\xi_{\rm line} L_{\rm disc}}{8 \pi c}
        \frac{2}{r^2} 
      = \frac{1}{\Gamma^2 (1 + \Beta)^2} 
        \frac{\xi_{\rm line} L_{\rm disc}}{4 \pi c r^2}.
\end{equation}
We have obtained Eq.~(\ref{eq:point_source_com}), again with $L_0 = \xi_{\rm line} L_{\rm disc}$.

\subsubsection{Ring dust torus}
Following \citet{finke_2016}, we model the DT as a geometrically thin ring of radius $R_{\rm DT}$. The incidence angle and the distance of the blob from the reprocessing material in Eq.~(\ref{eq:reprocessed_geom}) can be simplified to the ring case by imposing $\theta_{\rm re} = \pi / 2$ and fixing the radial coordinate to the ring radius: $r_{\rm re} = R_{\rm DT}$
\begin{equation}
\begin{split}
\mu_\ast &= \frac{r}{x}, \\
x^2 &= R^2_{\rm DT} + r^2.
\end{split}
\label{eq:repr_geom_ring}
\end{equation}
The specific spectral energy density for a ring DT reads \citep{finke_2016}
\begin{equation}
\phdensity(\epsilon, \Omega; r) = \frac{\xi_{\rm DT} L_{\rm disc}}{8 \pi^2 c x^2} 
    \delta\left(\mu - \frac{r}{x}\right) \delta(\epsilon - 2.7 \Theta),
\label{eq:u_dt}
\end{equation}
where $\xi_{\rm DT}$ is the fraction of the disc luminosity ($L_{\rm disc}$) reprocessed by the DT and we are approximating the torus black-body thermal emission as monochromatic at the peak energy $\epsilon_{\rm DT} = 2.7 \Theta$. The dimensionless temperature $\Theta = k_{\rm B} T_{\rm DT} / (m_{\rm e} c^2)$, with $k_{\rm B}$ Boltzmann constant and $T_{\rm DT}$ the peak temperature of the thermal emission. The integral energy density in the galactic frame is
\begin{equation}
u(r) = \frac{\xi_{\rm DT} L_{\rm disc}}{4 \pi c x^2}.
\label{eq:torus_stat}
\end{equation}
This can be straightforwardly reduced to Eq.~(\ref{eq:point_source_stat}) (with $L_0 = \xi_{\rm DT} L_{\rm disc}$) for $r \gg R_{\rm DT}$ (in which case $x^2 \rightarrow r^2$).
In the frame comoving with the blob the energy density reads
\begin{equation}
u'(r) = \Gamma^2 (1 - \Beta\,r/x)^2 \frac{\xi_{\rm DT} L_{\rm disc}}{4 \pi c x^2}.
\label{eq:torus_com}
\end{equation}
This expression can also be easily reduced to the case of the point-source behind the blob, Eq.~(\ref{eq:point_source_com}), considering that in the limit of large distances $\frac{x}{r} \rightarrow 1$ and $\Gamma^2 (1 - \Beta)^2 = 1 / [\Gamma^2 (1 + \Beta)^2]$.

\subsection{Synchrotron radiation}
\label{sec:synchrotron_radiation}
The assumptions used in the synchrotron radiation calculation are detailed in \ref{sec:processes}; here we report the formulae implemented in \agnpy.
\subsubsection{Spectral energy distribution}
The synchrotron SED is computed, following \citet{dermer_menon_2009} and \citet{finke_2008}, as
\begin{equation}
\hat{f}_{\hat{\epsilon}}^{\rm synch} = 
\frac{\delta_{\rm D}^4}{4 \pi d_{\rm L}^2} \, 
\epsilon' \frac{\sqrt{3} {\rm e}^3 B}{h} \int_{1}^{\infty} \diff\gamma'\;\enumber(\gamma')\,R(x),
\label{eq:synch_sed}
\end{equation}
where $d_{\rm L}$ is the luminosity distance of the source, $\epsilon'$ the energy of the emitted synchrotron photons in the blob reference frame, ${\rm e}$ the electron charge, and $B$ the magnetic field. $R(x)$ represents the pitch-angle-averaged synchrotron spectral power of a single electron, expressed as a function of the variable $x = (4 \pi \epsilon' m_{\rm e}^2 c^3)/(3 {\rm e} B h \gamma'^2)$. In \agnpy the approximation in Eq.~(D7) of \citet{aharonian_2010} is used for $R(x)$. 

\subsubsection{Synchrotron self-absorption}
The synchrotron self-absorption (SSA) effect \citep[see][Section 7.8]{dermer_menon_2009} is taken into account by mutliplying the SED in \ref{eq:synch_sed} by the absorption factor
\begin{equation}
\frac{3}{\tau_{\rm SSA}} \left( 
\frac{1}{2} + \frac{\exp(-\tau_{\rm SSA})}{\tau_{\rm SSA}} 
- \frac{1 - \exp(-\tau_{\rm SSA})}{\tau_{\rm SSA}^2} \right)
\label{eq:ssa_absorption}
\end{equation}
where the energy-dependent SSA opacity is
\begin{equation}
\tau_{\rm SSA}(\epsilon') = 
- \frac{R'_{\rm b}}{8 \pi m_{\rm e} \epsilon'^2} \left( \frac{\lambda_{\rm C}}{c} \right)^3 
\int_{1}^{\infty} \diff\gamma'\,R(x) 
\left[\gamma'^2 \frac{\partial}{\partial \gamma'} \left(\frac{\edensity}{\gamma'^2} \right) \right]
\label{eq:ssa_opacity}
\end{equation}
and where $\lambda_{\rm C} = h / (m_{\rm e} c)$ is the Compton wavelength.

\subsection{Inverse Compton radiation}
\label{sec:inverse_compton}
As for the synchrotron case, we refer to \ref{sec:processes} for the assumption used in the inverse Compton calculation and provide here the formulae implemented in \agnpy. Some detail is given here to mathematically formalise the assumption going in the external Compton calculations.

\subsubsection{Synchrotron self-Compton}
The synchrotron self-Compton SED is computed, following \citet{dermer_menon_2009} and \citet{finke_2008}, as
\begin{equation}
\begin{split}
\hat{f}_{\hat{\epsilon}_{\rm s}}^{\rm SSC} =& \frac{\delta_{\rm D}^4}{4 \pi d_{\rm L}^2} \, 
\frac{3}{4} c \sigma_{\rm T} \epsilon'^2_{\rm s} \,
        \int_0^\infty \diff\epsilon' \, \frac{\phdensity'_{\rm synch}(\epsilon')}{\epsilon'^2} \\
      & \int_0^\infty \diff\gamma \, \frac{\enumber(\gamma')}{\gamma'^2} \, F_{\rm C}(q', \Gamma'_{\rm e}),
\end{split}
\label{eq:ssc_sed}
\end{equation}
here $\sigma_{\rm T}$ is the Thomson cross section, $\epsilon'$ is the energy of the synchrotron photon, and $\epsilon'_{\rm s}$ the energy of the scattered photon, both in the blob reference frame. The specific spectral energy density of synchrotron photons $\phdensity'_{\rm synch}$ can be obtained from Eq.~(\ref{eq:synch_sed}) as
\begin{equation}
\phdensity'_{\rm synch} =  \frac{3}{4} \frac{3 d^2_{\rm L} \hat{f}_{\hat{\epsilon}}^{\rm synch}}{c R'^2_{\rm b} \delta_{\rm D}^4 \epsilon'}; \label{eq:phdens_synch}
\end{equation}
the factor $3/4$ is due to averaging the radiation in a sphere. $F_{\rm C}$ is an integration kernel representing the Compton cross section for electrons and photons with a uniform spatial distribution \citep[][]{jones_1968, blumenthal_gould_1970}
\begin{equation}
\begin{split}
F_{\rm C}(q', \Gamma'_{\rm e}) =& \bigg[ 2 q' \ln q' + (1 + 2 q')(1 - q') \bigg.\\
                                & +\left. \frac{1}{2}\frac{(\Gamma'_{\rm e} q')^2}{(1 + \Gamma'_{\rm e} q')}(1 - q') \right] 
                                H \left(q'; \frac{1}{4\gamma'^2}, 1 \right)\\
\end{split}
\label{eq:isotropic_kernel}
\end{equation}
with
\begin{equation}
\Gamma'_{\rm e} = 4 \epsilon'_{\rm s} \gamma' \;\;{\rm and}\;\; q' = \frac{\epsilon'_{\rm s}/\gamma'}{\Gamma'_{\rm e}(1 - \epsilon'_{\rm s}/\gamma')},
\end{equation}
and the Heaviside functions on $q'$ expresses kinematic limits on $\epsilon'$ and $\gamma'$. 

\subsubsection{External Compton}
In order to simplify the calculation of the Compton spectra for photon targets external to the blob, we employ the 'head-on' approximation \citep{reynolds_1982, dermer_1993, dermer_2009}: we assume that $\gamma' \gg 1$ and that, due to relativistic transformations of the angles, photons are scattered back approximately in the same direction as the incident electrons. Under this condition
\begin{equation}
\frac{\diff \sigma_{\rm C}}{\diff \epsilon_{\rm s} \diff \Omega_{\rm s}} \approx \frac{\diff \sigma_{\rm C}}{\diff \epsilon_{\rm s}} \delta(\Omega_{\rm s} - \Omega_{\rm e}), 
\label{eq:head_on_approx}    
\end{equation}
where $\epsilon_{\rm s}$ is the scattered photon energy, $\Omega_{\rm s}$ its direction that the delta ensures to be equal to the electron direction, $\Omega_{\rm e}$. To compute the Compton spectrum, the electron and photon densities have to be transformed to the same reference frame and then folded with the cross section. In \agnpy, we follow the approach proposed by \citet{gkm_2001} and convert the electron distribution to the galactic frame. For an isotropic electron distribution, $\underline{N}_{\rm e}(\gamma, \Omega_{\rm e}) = \delta_{\rm D}^3\,\enumber(\gamma/\delta_{\rm D})$. Given these assumptions, the energy flux produced by the blob electrons Compton-scattering an external radiation field reads 
\citep[]{dermer_menon_2009}
\begin{equation}
\begin{split}
\hat{f}_{\hat{\epsilon}_{\rm s}}^{\rm EC}
    =& \frac{c \pi r_{\rm e}^2}{4 \pi d_{\rm L}^2} \epsilon_{\rm s}^2 \delta_{\rm D}^3
    \int \diff\Omega \, \int_{0}^{\epsilon_{\rm high}} \diff\epsilon \, 
    \frac{\phdensity(\epsilon, \Omega; r)}{\epsilon^2} \\
    & \int_{\gamma_{\rm low}}^{\infty} \diff\gamma \, \frac{\enumber(\gamma / \delta_{\rm D})}{\gamma^2}\,\Xi_{\rm C},  
\end{split}
\label{eq:ec_sed}
\end{equation}
where $r_{\rm e}$ is the electron radius, $\epsilon_{\rm s}$ the dimensionless energy of the scattered photon in the galactic frame and $\phdensity$ the specific spectral energy density of the considered photon field. The Compton cross-section in the head-on approximation reduces to the integration kernel 
\begin{equation}
\begin{split}
\Xi_{\rm C} =& y + y^{-1} + \frac{2 \epsilon_{\rm s}}{\gamma \bar{\epsilon} y} 
              + \left( \frac{\epsilon_{\rm s}}{\gamma \bar{\epsilon} y} \right)^2, \\
{\rm with}\;y =& 1 - \frac{\epsilon_{\rm s}}{\gamma}, \\
{\rm and}\;\bar{\epsilon} =& \epsilon \gamma (1 - \cos\psi), \\
\end{split}
\label{eq:compton_kernel}
\end{equation}
where $\cos\psi$ is the angle between the direction of the incident photon and the scattering electron (see Fig.~\ref{fig:target_geometry})
\begin{equation}
\cos \psi = \mu \mu_{\rm s} + \sqrt{1 - \mu^2} \sqrt{1 - \mu_{\rm s}^2} \cos \phi.
\label{eq:cos_psi}
\end{equation}
The extremes of integration on the target energy $\epsilon_{\rm high}$ and Lorentz factor of the electron $\gamma_{\rm low}$ are imposed by kinematic limits
\begin{equation}
\begin{split}
\epsilon_{\rm high} &= \frac{2 \epsilon_{\rm s}}{1 - \cos \psi}, \\
\gamma_{\rm low} &= \frac{\epsilon_{\rm s}}{2} \left[ 1 + \sqrt{1 + \frac{2}{\epsilon \epsilon_{\rm s} (1 - \cos \psi)}} \right].
\end{split}
\label{eq:ec_kinematics}
\end{equation}
\par
Considering the specific spectral energy densities $\phdensity$ given in Sect.~\ref{sec:energy_densities}, we compute the Compton-scattered spectra corresponding to each of the targets implemented in \agnpy.
\par
If the target for external Compton scattering is an isotropic monochromatic radiation field, then plugging Eq.~(\ref{eq:u_iso}) in (\ref{eq:ec_sed}), we obtain
\begin{equation}
\begin{split}
\hat{f}_{\hat{\epsilon}_{\rm s}}^{\rm EC} =& 
    \frac{3}{2^7 \pi^2} \frac{c \sigma_{\rm T} u_0}{d_{\rm L}^2}  
    \left( \frac{\epsilon_{\rm s}}{\epsilon_0} \right)^2 \delta_{\rm D}^3 
    \int_0^{2 \pi} \diff\phi \, 
    \int_{-1}^{1} \diff\mu \, \\
    & \int_{\gamma_{\rm low}}^{\infty} \diff\gamma \, 
    \frac{\enumber(\gamma / \delta_{\rm D})}{\gamma^2}\,\Xi_{\rm C}. \\
\end{split}
\label{eq:ec_iso_sed}
\end{equation}
For a point source behind the blob, $\phdensity$ is given by Eq.~(\ref{eq:u_ps_behind_jet}), and the Compton-scattered SED reads
\begin{equation}
\hat{f}_{\hat{\epsilon}_{\rm s}}^{\rm EC}(r) = 
    \frac{3}{2^7 \pi^2} \frac{\sigma_{\rm T} L_0}{d_{\rm L}^2 r^2}  
    \left( \frac{\epsilon_{\rm s}}{\epsilon_0} \right)^2 \delta_{\rm D}^3 
    \int_{\gamma_{\rm low}}^{\infty} \diff\gamma \, \frac{\enumber(\gamma / \delta_{\rm D})}{\gamma^2}\,\Xi_{\rm C},
\label{eq:ec_ps_behind_jet_sed}
\end{equation}
where $r$ indicates the distance of the blob from the central BH (Fig.~\ref{fig:target_geometry}) and we remark the dependence of the SED on this quantity in the left side of the equation. We note that in this case $\mu=1 \Rightarrow \cos\psi = \mu_{\rm s}$. Considering a Shakura-Sunyaev accretion disc as a target, $\phdensity$ is Eq.~(\ref{eq:u_ss_disc}), one derives
\begin{equation}
\begin{split}
\hat{f}_{\hat{\epsilon}_{\rm s}}^{\rm EC}(r) =&
    \frac{3^2}{2^9 \pi^3}
    \frac{\sigma_{\rm T} G M \dot{m}}{d_{\rm L}^2 r^3} 
    \epsilon_{\rm s}^2 \delta_{\rm D}^3
    \int_0^{2 \pi} \diff\phi \, \\
    & \int_{\mu_{\rm min}}^{\mu_{\rm max}} \diff\mu \,
    \frac{\varphi(\mu; r)}{\epsilon^2_0(\mu; r) (\mu^{-2} - 1)^{3/2}} \\ 
    & \int_{\gamma_{\rm low}}^{\infty} \diff\gamma \,
    \frac{\enumber(\gamma / \delta_{\rm D})}{\gamma^2}\,\Xi_{\rm C}, \\
\end{split}
\label{eq:ec_ss_disc_sed}
\end{equation}
where the extremes of integration on the cosine of the zenith are given in Eq.~(\ref{eq:mu_limits_ss_disc}) and once more we have made explicit that the photons emitted at a given disc radius $R$ will have different incidence angles, depending on the blob position $r$, by replacing $\varphi(R) \rightarrow \varphi(\mu; r)$ and $\epsilon_0(R) \rightarrow \epsilon_0(\mu; r)$. The Compton kernel $\Xi_{\rm C}$ is dependent on $\mu$ through $\cos\psi$ and through the target energy $\epsilon = \epsilon_0(\mu; r)$. The external Compton spectrum for a spherical shell BLR target photon field, $\phdensity$ in Eq.~(\ref{eq:u_blr}), is
\begin{equation}
\begin{split}
\hat{f}_{\hat{\epsilon}_{\rm s}}^{\rm EC}(r) =& 
    \frac{3}{2^9 \pi^3} 
    \frac{\sigma_{\rm T} \xi_{\rm line} L_{\rm disc}}{d_{\rm L}^2}
    \left( \frac{\epsilon_{\rm s}}{\epsilon_{\rm line}} \right)^2
    \delta_{\rm D}^3 \int_0^{2 \pi} \diff\phi \,
    \int_{-1}^{1}\frac{\diff\mu_{\rm re}}{x^2} \, \\ 
    & \int_{\gamma_{\rm low}}^{\infty} \diff\gamma \, 
    \frac{\enumber(\gamma / \delta_{\rm D})}{\gamma^2}\,\Xi_{\rm C},
\end{split}
\label{eq:ec_blr_sed}
\end{equation}
where now the angle between the scattering electron and the target photon is  $\cos\psi = \mu_\ast \mu_{\rm s} + \sqrt{1 - \mu_\ast^2} \sqrt{1 - \mu_{\rm s}^2} \cos \phi$ with $\mu_\ast$ depending on the zenith angle of the reprocessing material $\mu_{\rm re}$ according to Eq.~(\ref{eq:reprocessed_geom}). Last, considering the photon field of a ring dust torus, $\phdensity$ in Eq.~(\ref{eq:u_dt}), we obtain the Compton-scattered SED
\begin{equation}
\begin{split}
\hat{f}_{\hat{\epsilon}_{\rm s}}^{\rm EC}(r) =&
    \frac{3}{2^8 \pi^3} \frac{\sigma_{\rm T} \xi_{\rm DT} L_{\rm disc}}{d_{\rm L}^2 x^2}
    \left( \frac{\epsilon_{\rm s}}{\epsilon_{\rm DT}} \right)^2 \delta_{\rm D}^3 
    \int_0^{2 \pi} \diff\phi \, \\
    & \int_{\gamma_{\rm low}}^{\infty} \diff\gamma \, 
    \frac{\enumber(\gamma / \delta_{\rm D})}{\gamma^2}\,\Xi_{\rm C}, \\
\end{split}
\label{eq:ec_dt_sed}
\end{equation}
where now $\cos\psi = \frac{r}{x} \mu_{\rm s} + \sqrt{1 - \left(\frac{r}{x}\right)^2} \sqrt{1 - \mu_{\rm s}^2} \cos \phi$.

\subsection{$\gamma\gamma$ absorption}
\label{sec:absorption}
The optical depth for $\gamma\gamma$ absorption for a photon with observed frequency $\hat{\nu}_1$ on a target photon field with specific spectral energy density $\phdensity(\epsilon, \Omega; r)$ is
\begin{equation}
\begin{split}
\tau_{\gamma \gamma}(\hat{\nu}_1) =& 
    \int_{0}^{\infty} \diff l \,
    \int_{0}^{2\pi} \diff\phi \, 
    \int_{-1}^{1}  \diff\mu \, (1 - \cos\psi) \, \\ 
    & \int_{0}^{\infty} \diff\epsilon \, 
    \frac{\phdensity(\epsilon, \Omega; l)}{\epsilon m_{\rm e} c^2} \, 
    \sigma_{\gamma \gamma}(s),\\
\end{split}
\label{eq:tau_gamma_gamma}
\end{equation}
where $\cos\psi$ is the cosine of the angle between the emitted and the absorbing photon (same as in Eq.~\ref{eq:cos_psi}), $\phi$ and $\mu$ the polar angles over which we integrate, and $l$ represents the distance from the emission region to the observer along which the absorption takes place. $\sigma_{\gamma \gamma}(s)$ represents the $\gamma\gamma$ absorption cross section
\begin{equation}
\begin{split}
\sigma_{\gamma \gamma}(s) &= \frac{1}{2} \pi r^2_{\rm e} (1 - \beta^2_{\rm cm}) 
\left[ (3 - \beta^4_{\rm cm}) \ln \left( \frac{1 + \beta_{\rm cm}}{1 - \beta_{\rm cm}} \right) \right. \\
& -2 \beta_{\rm cm} (2 -  \beta^2_{\rm cm}) \bigg] \\
\end{split}
\label{eq:sigma_gamma_gamma}
\end{equation}
where $r_{\rm e}$ is the electron radius, $\beta_{\rm cm} = \sqrt{1 - s^{-1}}$ and $s = \epsilon_1 \epsilon \, (1 - \cos\psi)\,/\,2$ are kinematic variables representing the momentum of the produced ${\rm e}^{\pm}$ pair in the centre-of-momentum frame. $\epsilon_1$ is the dimensionless energy of the photon emitted in the galactic frame (whose absorption we want to estimate), $\epsilon$ the dimensionless energy of the target photon. Both in \citet{dermer_2009} and \citet{finke_2016} the integral in Eq.~(\ref{eq:tau_gamma_gamma}) is simplified assuming that the emitted photons travel in the direction parallel to the jet axis ($\mu_{\rm s} \rightarrow 1 \Rightarrow \cos\psi = \mu$), decoupling the cross section and the $(1 - \cos\psi)$ term from the integral on $\phi$. The optical depths thus calculated are valid only for blazars (characterised by a small viewing angle to the observer line of sight). \agnpy carries out the full integration, computing optical depths valid for any jetted AGN. For speed reasons, \agnpy falls back to the simplified calculations in \citet{dermer_2009} and \citet{finke_2016} in the case in which $\mu_{\rm s} = 1$ is specified.

\subsubsection{Absorption on targets}

\begin{figure}
\centering
  
\tikzset {_1d93tyxgb/.code = {\pgfsetadditionalshadetransform{ \pgftransformshift{\pgfpoint{-198 bp } { -198 bp }  }  \pgftransformscale{1.32 }  }}}
\pgfdeclareradialshading{_jrroba7e8}{\pgfpoint{160bp}{160bp}}{rgb(0bp)=(1,1,1);
rgb(0bp)=(1,1,1);
rgb(25bp)=(0.48,0.15,0.15);
rgb(400bp)=(0.48,0.15,0.15)}

  
\tikzset {_tnbd2ijzd/.code = {\pgfsetadditionalshadetransform{ \pgftransformshift{\pgfpoint{0 bp } { 0 bp }  }  \pgftransformscale{1 }  }}}
\pgfdeclareradialshading{_bv0goexz0}{\pgfpoint{0bp}{0bp}}{rgb(0bp)=(1,1,1);
rgb(0bp)=(1,1,1);
rgb(25bp)=(0,0,0);
rgb(400bp)=(0,0,0)}
\tikzset{every picture/.style={line width=0.75pt}} 

\begin{tikzpicture}[x=0.75pt,y=0.75pt,yscale=-1,xscale=1]

\draw    (302.32,269.7) -- (495.95,269.7) ;
\draw [shift={(497.95,269.7)}, rotate = 180] [color={rgb, 255:red, 0; green, 0; blue, 0 }  ][line width=0.75]    (10.93,-3.29) .. controls (6.95,-1.4) and (3.31,-0.3) .. (0,0) .. controls (3.31,0.3) and (6.95,1.4) .. (10.93,3.29)   ;
\draw    (302.32,269.7) -- (176.46,344.37) ;
\draw [shift={(174.74,345.39)}, rotate = 329.32] [color={rgb, 255:red, 0; green, 0; blue, 0 }  ][line width=0.75]    (10.93,-3.29) .. controls (6.95,-1.4) and (3.31,-0.3) .. (0,0) .. controls (3.31,0.3) and (6.95,1.4) .. (10.93,3.29)   ;
\draw    (302.32,269.7) -- (301.55,21.55) ;
\draw [shift={(301.54,19.55)}, rotate = 449.82] [color={rgb, 255:red, 0; green, 0; blue, 0 }  ][line width=0.75]    (10.93,-3.29) .. controls (6.95,-1.4) and (3.31,-0.3) .. (0,0) .. controls (3.31,0.3) and (6.95,1.4) .. (10.93,3.29)   ;
\draw [color={rgb, 255:red, 74; green, 144; blue, 226 }  ,draw opacity=1 ][line width=1.5]  [dash pattern={on 5.63pt off 4.5pt}]  (302.32,269.7) -- (365.88,293.37) ;
\draw [color={rgb, 255:red, 74; green, 144; blue, 226 }  ,draw opacity=1 ][line width=1.5]    (288.45,278.57) .. controls (300.01,284.07) and (311.56,283.28) .. (321.57,277) ;
\draw [color={rgb, 255:red, 208; green, 2; blue, 27 }  ,draw opacity=1 ][line width=1.5]    (302.32,94.76) -- (251.47,36.26) ;
\draw [shift={(249.5,34)}, rotate = 409] [color={rgb, 255:red, 208; green, 2; blue, 27 }  ,draw opacity=1 ][line width=1.5]    (14.21,-4.28) .. controls (9.04,-1.82) and (4.3,-0.39) .. (0,0) .. controls (4.3,0.39) and (9.04,1.82) .. (14.21,4.28)   ;
\path  [shading=_jrroba7e8,_1d93tyxgb] (295.35,94.76) .. controls (295.35,90.81) and (298.47,87.61) .. (302.32,87.61) .. controls (306.18,87.61) and (309.3,90.81) .. (309.3,94.76) .. controls (309.3,98.71) and (306.18,101.91) .. (302.32,101.91) .. controls (298.47,101.91) and (295.35,98.71) .. (295.35,94.76) -- cycle ; 
 \draw  [color={rgb, 255:red, 0; green, 0; blue, 0 }  ,draw opacity=1 ] (295.35,94.76) .. controls (295.35,90.81) and (298.47,87.61) .. (302.32,87.61) .. controls (306.18,87.61) and (309.3,90.81) .. (309.3,94.76) .. controls (309.3,98.71) and (306.18,101.91) .. (302.32,101.91) .. controls (298.47,101.91) and (295.35,98.71) .. (295.35,94.76) -- cycle ; 

\draw [color={rgb, 255:red, 208; green, 2; blue, 27 }  ,draw opacity=1 ][line width=1.5]    (302.01,47.62) .. controls (285.5,51) and (286,52) .. (275.91,64.38) ;
\draw  [color={rgb, 255:red, 74; green, 144; blue, 226 }  ,draw opacity=1 ][dash pattern={on 5.63pt off 4.5pt}][line width=1.5]  (198.63,269.7) .. controls (198.63,210.99) and (245.06,163.4) .. (302.32,163.4) .. controls (359.59,163.4) and (406.02,210.99) .. (406.02,269.7) .. controls (406.02,328.41) and (359.59,376) .. (302.32,376) .. controls (245.06,376) and (198.63,328.41) .. (198.63,269.7) -- cycle ;
\draw [color={rgb, 255:red, 74; green, 144; blue, 226 }  ,draw opacity=1 ][line width=1.5]    (302.32,269.7) -- (364.46,216.11) ;
\draw [shift={(366.73,214.15)}, rotate = 499.22] [color={rgb, 255:red, 74; green, 144; blue, 226 }  ,draw opacity=1 ][line width=1.5]    (14.21,-4.28) .. controls (9.04,-1.82) and (4.3,-0.39) .. (0,0) .. controls (4.3,0.39) and (9.04,1.82) .. (14.21,4.28)   ;
\draw [color={rgb, 255:red, 74; green, 144; blue, 226 }  ,draw opacity=1 ] [dash pattern={on 0.84pt off 2.51pt}]  (366.73,214.15) -- (365.88,293.37) ;
\draw [color={rgb, 255:red, 74; green, 144; blue, 226 }  ,draw opacity=1 ][line width=1.5]    (302.61,209.71) .. controls (317.64,211.91) and (334.12,219.19) .. (344.7,232.04) ;
\draw  [color={rgb, 255:red, 74; green, 144; blue, 226 }  ,draw opacity=1 ][dash pattern={on 1.69pt off 2.76pt}][line width=1.5]  (197.97,269.7) .. controls (197.97,253.47) and (244.69,240.31) .. (302.32,240.31) .. controls (359.96,240.31) and (406.68,253.47) .. (406.68,269.7) .. controls (406.68,285.93) and (359.96,299.09) .. (302.32,299.09) .. controls (244.69,299.09) and (197.97,285.93) .. (197.97,269.7) -- cycle ;
\draw [color={rgb, 255:red, 65; green, 117; blue, 5 }  ,draw opacity=1 ][line width=1.5]    (302.32,269.7) -- (302.32,97.76) ;
\draw [shift={(302.32,94.76)}, rotate = 450] [color={rgb, 255:red, 65; green, 117; blue, 5 }  ,draw opacity=1 ][line width=1.5]    (14.21,-4.28) .. controls (9.04,-1.82) and (4.3,-0.39) .. (0,0) .. controls (4.3,0.39) and (9.04,1.82) .. (14.21,4.28)   ;
\path  [shading=_bv0goexz0,_tnbd2ijzd] (295.35,269.7) .. controls (295.35,265.75) and (298.47,262.55) .. (302.32,262.55) .. controls (306.18,262.55) and (309.3,265.75) .. (309.3,269.7) .. controls (309.3,273.65) and (306.18,276.85) .. (302.32,276.85) .. controls (298.47,276.85) and (295.35,273.65) .. (295.35,269.7) -- cycle ; 
 \draw   (295.35,269.7) .. controls (295.35,265.75) and (298.47,262.55) .. (302.32,262.55) .. controls (306.18,262.55) and (309.3,265.75) .. (309.3,269.7) .. controls (309.3,273.65) and (306.18,276.85) .. (302.32,276.85) .. controls (298.47,276.85) and (295.35,273.65) .. (295.35,269.7) -- cycle ; 

\draw [line width=1.5]    (251.14,36.51) -- (366.73,214.15) ;
\draw [shift={(249.5,34)}, rotate = 56.95] [color={rgb, 255:red, 0; green, 0; blue, 0 }  ][line width=1.5]    (14.21,-4.28) .. controls (9.04,-1.82) and (4.3,-0.39) .. (0,0) .. controls (4.3,0.39) and (9.04,1.82) .. (14.21,4.28)   ;

\draw (340.65,258) node  [color={rgb, 255:red, 74; green, 144; blue, 226 }  ,opacity=1 ]  {$\mathbf{r}_{\mathrm{re}}$};
\draw (292.51,61.72) node  [color={rgb, 255:red, 208; green, 2; blue, 27 }  ,opacity=1 ]  {$\theta _{\mathrm{s}}$};
\draw (238.69,23.2) node   [align=left] {\textcolor[rgb]{0.82,0.01,0.11}{observer}};
\draw (330.76,90.94) node   [align=left] {blob};
\draw (287.09,258.85) node   [align=left] {BH};
\draw (330.25,202.58) node  [color={rgb, 255:red, 74; green, 144; blue, 226 }  ,opacity=1 ]  {$\theta _{\mathrm{re}}$};
\draw (455.95,243.45) node  [color={rgb, 255:red, 74; green, 144; blue, 226 }  ,opacity=1 ] [align=left] {reprocessing\\material};
\draw (292.31,286.52) node  [color={rgb, 255:red, 74; green, 144; blue, 226 }  ,opacity=1 ]  {$\phi _{\mathrm{re}}$};
\draw (322.22,123.07) node  [color={rgb, 255:red, 74; green, 144; blue, 226 }  ,opacity=1 ]  {$\textcolor[rgb]{0,0,0}{\mathbf{x}}$};
\draw (274.09,43.57) node  [color={rgb, 255:red, 74; green, 144; blue, 226 }  ,opacity=1 ]  {$\textcolor[rgb]{0.82,0.01,0.11}{\mathbf{l}}$};
\draw (294.9,144.36) node  [color={rgb, 255:red, 155; green, 155; blue, 155 }  ,opacity=1 ]  {$\textcolor[rgb]{0.25,0.46,0.02}{\mathbf{r}}$};
\draw (176.39,351.21) node [anchor=north west][inner sep=0.75pt]    {$x$};
\draw (476.11,279.34) node [anchor=north west][inner sep=0.75pt]    {$y$};
\draw (314,15.7) node [anchor=north west][inner sep=0.75pt]    {$z$};

\end{tikzpicture}
\caption{Geometry used for the absorption calculation in case of arbitrary viewing angle $\theta_s$.}
\label{fig:target_geometry_arbitrary_mu_s}
\end{figure}
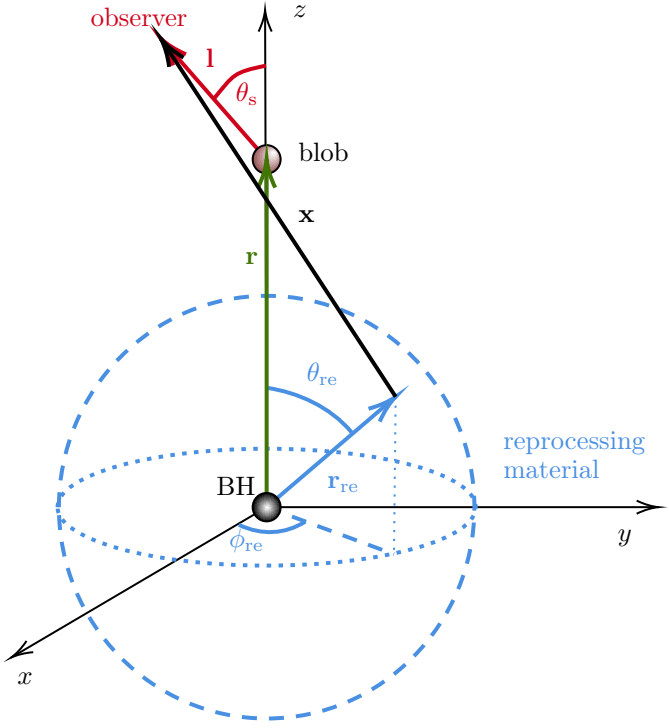

Figure~\ref{fig:target_geometry_arbitrary_mu_s} illustrates the geometry used for the opacity calculation in case of sources misaligned with the observer, those whose jet axis forms an angle $\theta_{\rm s}$ to the line of sight. In Eq.~(\ref{eq:tau_gamma_gamma}) the distance over which we are integrating runs along the line of sight and it is marked by the vector $\mathbf{l}$ in the figure. We can use simple vector algebra to relate it to the distance from the reprocessing material, $\mathbf{x}$, and the distance of the blob from the BH centre, $\mathbf{r}$. Assuming that the line of sight lies in the $x-z$ plane, we can write
\begin{equation}
    \begin{split}
        \mathbf{x} &= \mathbf{r} + \mathbf{l} - \mathbf{r}_{\rm re}\\
        \mathbf{r} &= (0, 0, r)\\
        \mathbf{l} &= (l \sin\theta_s, 0, l \cos\theta_s)\\
        \mathbf{r}_{\rm re} &= (
            r_{\rm re}\sin\theta_{\rm re}\cos\phi_{\rm re}, 
            r_{\rm re}\sin\theta_{\rm re}\sin\phi_{\rm re}, 
            r_{\rm re}\cos\theta_{\rm re}),\\
    \end{split}
\end{equation}
where we have used the same spherical coordinates of Fig.~\ref{fig:reprocessing_target_geometry} $(r_{\rm re}, \theta_{\rm re}, \phi_{\rm re})$ to characterise the reprocessing material.
The distance between a point on the reprocessing surface and a point on the line of sight $l$ is
\begin{equation}
    \mathbf{x} = \begin{pmatrix}
        l \sin\theta_s - r_{\rm re}\sin\theta_{\rm re}\cos\phi_{\rm re}\\
        - r_{\rm re}\sin\theta_{\rm re}\sin\phi_{\rm re}\\
        r + l \cos\theta_s - r_{\rm re}\cos\theta_{\rm re}
    \end{pmatrix}
\end{equation}
and its magnitude
\begin{equation}
    \begin{split}
    x^2 &= l^2 + r_{\rm re}^2 + r^2 
        - 2 l r_{\rm re} \sin\theta_{\rm re}\sin\theta_{\rm s} \cos\phi_{\rm re}\\  
        &- 2 r_{\rm re} \cos\theta_{\rm re}(r + l\mu_{\rm s}) 
        + 2 r l \mu_{\rm s}.
    \end{split}
\label{eq:reprocessed_geom_arbitrary_mu_s}
\end{equation}
Assuming $\mu_{\rm s} = 1$, at the blob position, $l=0$, Eq.~(\ref{eq:reprocessed_geom_arbitrary_mu_s}) reduces to the distance calculation employed for the EC in case of reprocessing material, Eq.~(\ref{eq:reprocessed_geom}).
\par
Considering a point-like source behind the jet and inserting its specific spectral energy density, Eq.~(\ref{eq:u_ps_behind_jet}), in Eq.~(\ref{eq:tau_gamma_gamma}), we obtain
\begin{equation}
\tau_{\gamma\gamma} = \frac{L_0}{4 \pi \epsilon_0 m_{\rm e} c^2} \int_{0}^{\infty}\diff l\,\frac{(1 - \cos\psi) \sigma_{\gamma\gamma}(s)}{x^2},
\label{eq:tau_ps_behind_jet}
\end{equation}
where the cosine of the angle between the photons is $\mu = (r + l \mu_{\rm s}) / x$. 
\par
Considering a spherical shell BLR and replacing its specific spectral energy density, Eq.~\ref{eq:u_blr}, in the general opacity formula, Eq.~(\ref{eq:tau_gamma_gamma}), we obtain
\begin{equation}
\begin{split}
\tau_{\gamma \gamma}(\hat{\nu}_1) &=
    \frac{1}{4 \pi^2} \frac{\xi_{\rm line} L_{\rm disc}}{\epsilon_{\rm line} m_{\rm e} c^3} \,
    \int_{0}^{\infty} \diff l \,
    \int_{0}^{2\pi} \diff\phi_{\rm re} \, \\
    & \int_{-1}^{1}\frac{\diff\mu_{\rm re}}{x^2} \, (1 - \cos\psi) \,
    \sigma_{\gamma \gamma}(s). \\
\label{eq:tau_blr}
\end{split}
\end{equation}
In this case, we use for the distance $x$ Eq.~(\ref{eq:reprocessed_geom_arbitrary_mu_s}) with $r_{\rm re} = R_{\rm line}$. The angles indicating the photon direction are $\mu = x_z / x$ and $\phi = \arctan\left(\frac{x_y}{x_x}\right)$. In the case $\mu_{\rm s} = 1$, one can see from Fig.~\ref{fig:target_geometry_arbitrary_mu_s} that $\vec{l}$ lies on the $z$ axis and the integration along the line of sight can be carried out along the $z$ axis, from $r$ to $\infty$ \citep[][Sec.~7]{finke_2016}. The integration over $\phi$ becomes trivial. 
\par
For the case of the DT, replacing its specific spectral energy density, Eq.~(\ref{eq:u_dt}), in the general opacity formula, Eq.~(\ref{eq:tau_gamma_gamma}), we obtain
\begin{equation}
\tau_{\gamma \gamma}(\hat{\nu}_1) =
    \frac{1}{8 \pi^2} \frac{\xi_{\rm DT} L_{\rm disc}}{\epsilon_{\rm DT} m_{\rm e} c^3} \,
    \int_{0}^{\infty} \diff l \,
    \int_{0}^{2\pi} \diff\phi \, \frac{(1 - \cos\psi)}{x^2} \,
    \sigma_{\gamma \gamma}(s).
\label{eq:tau_dt}
\end{equation}
The distance $x$ is obtained from Eq.~(\ref{eq:reprocessed_geom_arbitrary_mu_s}) with $r_{\rm re} = R_{\rm DT}$ and $\sin\theta_{\rm re} = 0$. The angles indicating the photon direction can be obtained as remarked for the BLR. Considering the case $\mu_{\rm s} = 1$, the same simplifications described in the BLR case can be applied. Photoabsorption by fields external to the emission region results in an attenuation of the photon flux by a factor $\exp(-\tau_{\gamma \gamma})$.

\subsubsection{Absorption on synchrotron radiation}
\agnpy includes also a simplified calculations of optical depth of high energy photons on the synchrotron radiation field using an approach similar to the one in \citet{finke_2008}, that is without taking into account the inhomogeneity and anisotropy of the radiation field. Contrary to \citet{finke_2008}, however, we assume that the total distance over which the absorption can occur is equal to the diameter (rather than radius) of the blob, and that the two photons collide at the angle of $\pi/2$. The optical depth is then computed (in the blob frame) as:
\begin{equation}
  \tau_{\rm synch}(\epsilon'_1) = 2 R'_b 
  \int_{\epsilon'_{\min}}^{\epsilon'_{\max}} \diff\epsilon \, 
  \frac{\phdensity'_{\rm synch}}{\epsilon' m_{\rm e} c^2}
  \sigma_{\gamma\gamma}(s),
\label{eq:tau_ssc}
\end{equation}
where $s=\epsilon'_1 \epsilon'/2$ and $\phdensity'_{\rm synch}$ is the differential energy density of the synchrotron photons averaged over the blob volume, following Eq.~(\ref{eq:phdens_synch}). It should be noticed that both the gamma-ray emission in the blob and the soft radiation on which it is absorbed fill homogeneously the blob. Therefore the attenuation factor, instead of $\exp(-\tau)$, which is the case for the attenuation of the blob emission in large-scale external radiation fields, is computed as \citep[see e.g.][]{finke_2008}
\begin{equation}
\frac{1-\exp(-\tau_{\rm synch})}{\tau_{\rm synch}}. 
\label{eq:abs_homo}
\end{equation}

\section{Physical parameters for the examples and validation}
The physical parameters used in all the examples and validations provided in the paper are listed in Tables~\ref{tab:blob_parameters} and \ref{tab:target_parameters}. Table~\ref{tab:blob_parameters} provides the parameters of the emission regions and their EEDs, Table~\ref{tab:target_parameters} provides instead those of the line and thermal emitters.

\begin{table}
\begin{minipage}{\linewidth}
\caption{Parameters of the blobs and the corresponding EEDs used in the examples and validations. \protect\footnote{For the SSC and EC model the same emission regions considered in Fig.~7.4 of \citet{dermer_menon_2009} and Table~1 of \citet{finke_2016}, respectively, were adopted.}}
\label{tab:blob_parameters}    
\centering     
\begin{tabular}{lll}
\hline\hline
quantity & SSC model & EC model \\
\hline
$W'_{\rm e}\,/\,{\rm erg}$ &  $10^{48}$ & $6 \times 10^{42}$ \\ 
$p$ & $2.8$ & \multicolumn{1}{l}{--} \\ 
$p_1$ & \multicolumn{1}{l}{--} & $2.0$ \\ 
$p_2$ & \multicolumn{1}{l}{--} & $3.5$ \\ 
$\gamma'_{\rm b}$ & \multicolumn{1}{l}{--} & $10^4$ \\
$\gamma'_{\min}$ & $10^2$ & $20$ \\
$\gamma'_{\max}$ & $10^7$ & $5\times 10^7$ \\
$R'_{\rm b}\,/\,{\rm cm}$ & $10^{16}$ & $10^{16}$ \\
$B\,/\,{\rm G}$ & $1$ & $0.56$ \\
$\Gamma$ & $10$ & $40$ \\
$\delta_{\rm D}$ & $10$ & $40$ \\
$z$ & $0.07$ & $1$ \\
\hline
\hline
\end{tabular}
\end{minipage}
\end{table}

\begin{table}
\begin{minipage}{\linewidth}
\caption{Parameters of the line and thermal emitters used in all the examples and validations. \protect\footnote{The same values in Table~1 of \citet{finke_2016} were adopted. Parameters of the ${\rm H\alpha}$-emitting BLR are obtained from the ${\rm Ly \alpha}$ ones with the scaling relations in the Appendix of \citet{finke_2016}. The radius of the dust torus is set to the sublimation radius in Eq.~(96) of \citet{finke_2016}.}}
\label{tab:target_parameters}      
\centering                                      
\begin{tabular}{ll}          
\hline\hline
quantity & value \\
\hline
$L_{\rm disc}\,/\,({\rm erg}\,{\rm s}^{-1})$ & $2 \times 10^{46}$ \\
$\eta$ & $1 / 12$ \\
$M_{\rm BH}\,/\,M_{\odot}$ &  $1.2 \times 10^9$ \\
$R_{\rm g}/{\rm cm}$ & $1.8 \times 10^{14}$ \\ 
$R_{\rm in}/R_{\rm g}$ & $6$ \\
$R_{\rm out}/R_{\rm g}$ & $200$ \\ 
\hline
$\xi_{\rm Ly\alpha}\,(\xi_{\rm H\alpha})$ & $0.024\,(0.007)$ \\
$\lambda_{\rm Ly\alpha}\,(\lambda_{\rm H\alpha}) /{\rm cm}$ & $1.22 \, (6.56) \times 10^{-5}$ \\
$R_{\rm Ly\alpha}\,(R_{\rm H\alpha}) \,/\,{\rm cm}$ & $1.1 \, (5.30) \times 10^{17}$ \\ 
\hline
$\xi_{\rm DT}$ & $0.1$ \\
$T_{\rm DT}/{\rm K}$ & $1000$ \\
$R_{\rm DT}/{\rm cm}$ & $1.57 \times 10^{19}$ \\ 
\hline
\hline
\end{tabular}
\end{minipage}
\end{table}

\section{Effect of the resolution of the discretisation on the smoothness of the spectra}
\label{sec:discretisation}

\begin{figure*}
\centering
    \includegraphics[width=17cm]{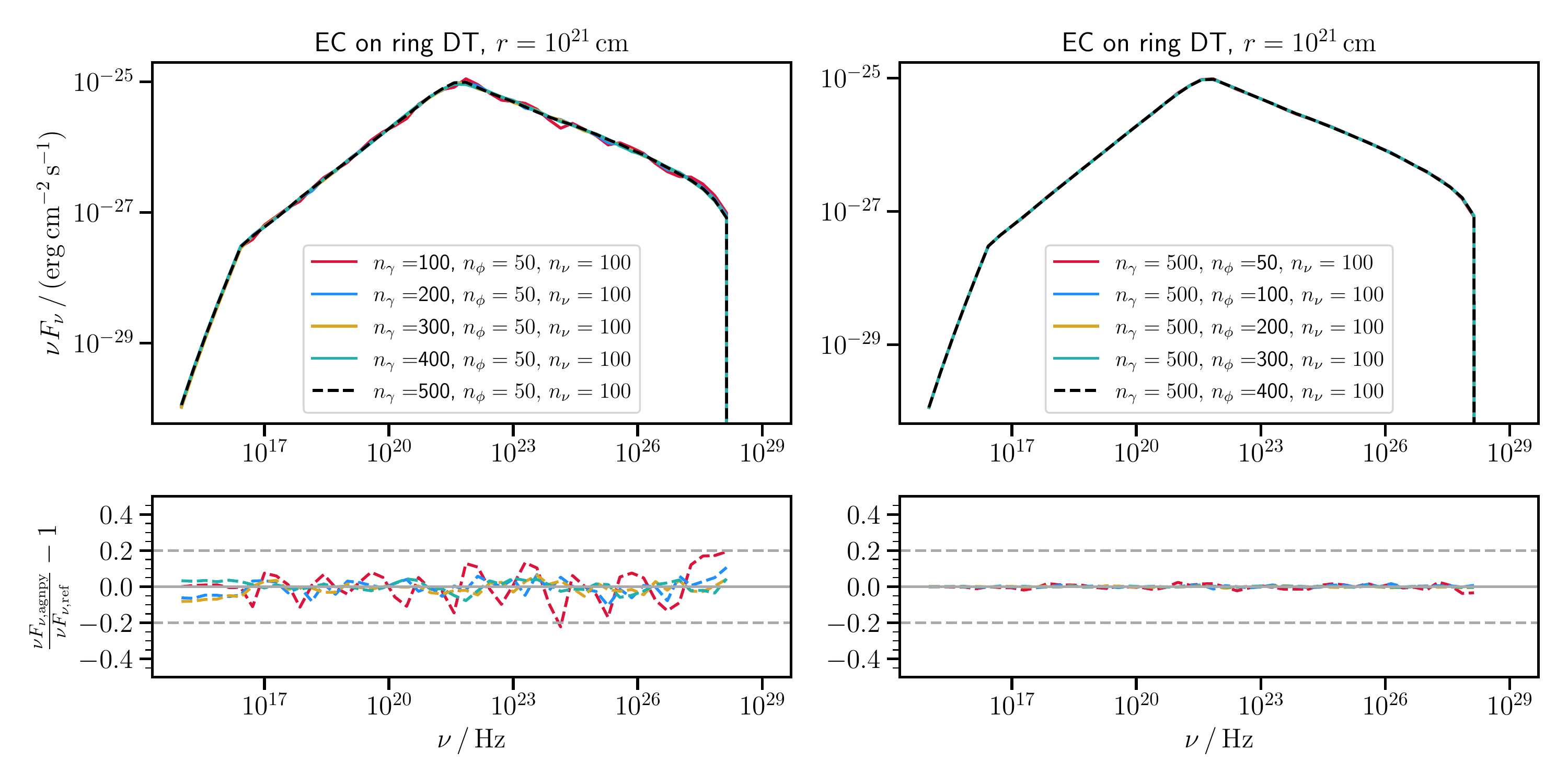}
    \caption{Effect of the resolution of the integration grid on the final spectrum. Left: variation of the number of points along the EED energy dimension, $n_{\gamma}$. Right: variation of the number of points along the azimuth angle dimension, $n_{\phi}$. In both bottom panels the deviation of the individual SEDs from the one with the denser integration grid (drawn with a dashed black line in the upper panel) is plotted.}
    \label{fig:sed_resolution}
\end{figure*}

In this section we briefly illustrate the effect of the discretisation of the integral on the smoothness of the final spectrum. We choose as a characteristic case the SED for EC on DT, as it contains a double integral with an energetic and a spatial dimensions represented by the EED energy and the azimuth angle, respectively (see Eq.~\ref{eq:ec_dt_sed}). Figure~\ref{fig:sed_resolution} shows the results of computing the SED with a fixed dimension of frequency values ($n_{\nu}=100$) and changing in one case (left panel) the number of points along the EED energy dimension, $n_{\gamma}$ logarithmically spaced values, and in the other (right panel) the number of points along the azimuth dimension, $n_{\phi}$ linearly spaced values. The bottom panels represent the deviations of the individual SEDs from the one with the denser integration grid (drawn in black). For this specific case it is evident that one can obtain a smoother spectrum by increasing the resolution along the EED energy dimension, while increasing the resolution along the azimuth dimension has no significant effect. We remark that the variation of the number of points along the spatial dimensions can have a different and more relevant impact if different parameters or a completely different physical process are considered. The ultimate resolution one can achieve depends on the machine used and its local memory, as larger integration grids will generate larger \numpy arrays in memory. Of course results might improve with the integration technique; we find the simple trapezoidal rule applied to logarithmically or linearly spaced integration grids acceptable for the processes described in this paper.

\end{appendix}

\end{document}